\documentclass[ijds,nonblindrev]{informs4}

\OneAndAHalfSpacedXI

\usepackage{array, amsmath, amssymb, amsfonts, hyperref, float,subcaption, listings}
\usepackage{tabularx}
\usepackage{caption}
\usepackage{subcaption}
\usepackage{tikz}
\usetikzlibrary{positioning}
\usetikzlibrary{arrows.meta}
\usepackage{lmodern} % Latin Modern font, a better version of Computer Modern
\usepackage{mathptmx} % Times font for both text and math

\usepackage[skip=2pt,font=large]{caption}

\usepackage[numbers,sort]{natbib}

\usepackage{tikz}
\usetikzlibrary{shapes.geometric, arrows.meta, positioning, fit}
\usepackage{hyperref}
\usepackage{xcolor}
\definecolor{darkred}{rgb}{0.55,0.0,0.0}
\definecolor{darkviolet}{rgb}{0.58, 0.0, 0.83}
\definecolor{darkblue}{rgb}{0.0, 0.0, 0.55}
\definecolor{linkblue}{RGB}{0, 102, 204}
\hypersetup{
    colorlinks=true,
    linkcolor=linkblue,
    citecolor=black,
    filecolor=linkblue,
    urlcolor=linkblue,
}
\usepackage{algorithm}
\usepackage[noend]{algpseudocode}
 \bibpunct[, ]{(}{)}{,}{a}{}{,}%
 \def\bibfont{\small}%
\usepackage{braket}
\usepackage{url}
\def\BibTeX{{\rm B\kern-.05em{\sc i\kern-.025em b}\kern-.08em
    T\kern-.1667em\lower.7ex\hbox{E}\kern-.125emX}}

\usepackage{rotating}
\usepackage{fancyvrb}
\lstdefinestyle{mystyle}{
    language=Python,
    basicstyle=\ttfamily,
    keywordstyle=\color{blue},
    commentstyle=\color{green},
    stringstyle=\color{black},
    breaklines=true,
    showstringspaces=false,
    numbers=left,
    numberstyle=\tiny,
    frame=single
}
\TheoremsNumberedThrough     % Preferred (Theorem 1, Lemma 1, Theorem 2)
\ECRepeatTheorems
\JOURNAL{2025 INFORMS TutORials}

\EquationsNumberedThrough    % Default: (1), (2), ...
\MANUSCRIPTNO{IJDS-0001-1922.65}

\begin{document}
%\setcounter{page}{1}
%\DOI{10.1287/xxxx.0000.0000}%


\RUNTITLE{Five Starter Problems: Solving Quadratic Unconstrained Binary Optimization Models on Quantum Computers}

\TITLE{Five Starter Problems: Solving Quadratic Unconstrained Binary Optimization Models on Quantum Computers}

\ARTICLEAUTHORS{%
\AUTHOR{Arul Rhik Mazumder}
% ,\textsuperscript{a} Second Author,\textsuperscript{b} Third Author,\textsuperscript{c} Fourth Author,\textsuperscript{c}

\AFF{Quantum Technologies Group, School of Computer Science, Carnegie Mellon University, Pittsburgh, PA 15213, USA, \EMAIL{arulm@andrew.cmu.edu}}
% \textsuperscript{b}School of Industrial Engineering, Good College, Collegeville, Maine 01234 \EMAIL{secauth@goodcoll.edu}; 
% \textsuperscript{c}Their Common Affiliation \EMAIL{thauth@anywhere.edu, fourauth@anywhere.edu}

\AUTHOR{Sridhar Tayur}

\AFF{Quantum Technologies Group, Tepper School of Business, Carnegie Mellon University, Pittsburgh, PA 15213, USA, \EMAIL{stayur@cmu.edu}}}

\ABSTRACT{%
% Several articles and books adequately cover quantum computing concepts, such as gate/circuit model, Quantum Approximate Optimization Algorithm (QAOA), Adiabatic Quantum Computing (AQC), and quantum annealing (QA). However, they typically stop short of accessing quantum hardware and only solve numerical problem instances.
This tutorial offers a quick, hands-on introduction to solving Quadratic Unconstrained Binary Optimization (QUBO) models on currently available quantum computers and their simulators. We cover both IBM and D-Wave machines: IBM utilizes a gate-circuit architecture, and D-Wave is a quantum annealer. We provide examples of three canonical problems % (Number Partitioning, Max-Cut, Minimum Vertex Cover) 
and two models from practical applications. % (from cancer genomics and a hedge fund portfolio manager)
The tutorial is structured to bridge the gap between theory and practice: we begin with an overview of QUBOs, explain their relevance and connection to quantum algorithms, introduce key quantum computing concepts, provide the foundations for two quantum heuristics, and provide detailed implementation guides. An associated GitHub repository provides the codes in five companion notebooks. In addition to reaching undergraduate and graduate students in computationally intensive disciplines, this article aims to reach working industry professionals seeking to explore the potential of near-term quantum applications. As our title indicates, this tutorial is intended to be a starting point in a journey towards solving more complex QUBOs on quantum computers. }

\maketitle

\section{Introduction}

Quantum computing is one of the most promising new technologies of the 21st century. However, due to the nascency of the hardware, many of the envisioned speedups remain largely theoretical. 
Despite limitations of available quantum hardware, there remains significant interest in exploring them. Although several articles and books (such as \citep{NielsenChuang11}) adequately cover the principles of quantum computing, they typically stop short of providing a road map to access quantum hardware and numerically solve problems of interest. This tutorial aims to provide quick access to IBM and D-Wave platforms, thus covering two types of quantum algorithms, one based on the gate-circuit model \citep{RP2014} and the other on quantum annealing \citep{CCM2014}, with {\em notebooks} containing associated software. We hope that our tutorial can serve as a companion to the available books and articles.

We first briefly introduce Quadratic Unconstrained Binary Optimization (QUBO). We then show how QUBOs model three canonical combinatorial optimization problems: Number Partitioning \citep{PhysRevLett.81.4281}, Max-Cut \citep{Commander2009}, and Minimum Vertex Cover \citep{10.1007/11821069_21}. We next model two practical problems: Order partitioning (for A / B testing in a hedge fund) and de novo discovery of driver (mutated) genes in the field of cancer genomics \citep{Alghassi845719}.
%Second, we will provide a brief primer on quantum computing and theoretical underpinnings for each of the key algorithms. 
We then provide step-by-step instructions\footnote{For completeness (and for ability to compare with a classical benchmark), we also include simulated annealing (SA) in our tutorial.} to solve QUBO models in two frameworks: the gate-circuit model (IBM hardware \citep{Qiskit}) and quantum annealing (D-Wave hardware \citep{rieffel2015case}).

Although numerous tutorials on quantum computing exist, including those offered by Qiskit \citep{Qiskit-Textbook} and D-Wave \citep{DWaveOceanDocs}, they often either focus narrowly on programming without contextual applications or emphasize theoretical aspects without concrete implementation workflows. Our tutorial fills this gap by providing a hands-on, data-driven pathway for newcomers to quantum computing who wish to go beyond textbook theory and actually run algorithms on quantum hardware. Unlike many existing resources, we present end-to-end examples, starting from formulating problems as QUBOs, mapping them to real-world datasets, and executing them using gate-based and annealing-based heuristics. As such, this tutorial serves as a practical companion to foundational texts such as \citep{NielsenChuang11} and is especially designed for readers who want to learn by doing. We aim to bridge the divide between abstract quantum concepts and their tangible, executable counterparts. We view this to be the first step in a journey towards solving more complex QUBOs on quantum computers.

The rest of the tutorial is structured as follows (see Figure~\ref{fig:flow-of-topics}). In Section~\ref{sec:qubo}, we introduce the QUBO model, which serves as the foundation for all subsequent applications and algorithmic developments. Section \ref{sec:problems} explores the classical problems that can be formulated as QUBOs, including number partitioning (Section \ref{subsec:np}), Max-Cut (Section \ref{subsec:mc}) and minimum vertex cover (Section \ref{subsec:mvc}). Section \ref{sec:practproblems} builds on these by highlighting advanced applications, such as order partitioning (Section \ref{subsec:op}) and cancer genomics (Section \ref{subsec:cg}), which demonstrate the versatility of QUBOs in real-world problem domains. Section \ref{sec:class-to-quantum} provides theoretical foundations: Part I (Section \ref{subsec:found1}) introduces the essential building blocks, and Part II (Section \ref{subsec:found2}) presents key algorithms to solve QUBOs, including simulated annealing (Section \ref{subsec:sa}), quantum annealing (Section \ref{subsec:qa}), and the quantum approximate optimization algorithm (QAOA) (Section \ref{subsec:qaoa}). Finally, Section \ref{sec:solve} focuses on implementations of these solvers in practice, with subsections dedicated to custom and open-source implementations of QAOA (Sections \ref{subsec:vqaoa},  \ref{subsec:oqaoa}, and \ref{subsec:qqaoa}), as well as implementations of simulated annealing (Section \ref{subsec:sai}) and quantum annealing (Section \ref{subsec:qai}).  We conclude in Section \ref{sec:conclusion}.

% \section{Mathematical Formulations for Optimization}
% This tutorial explores two key formulations for combinatorial optimization problems: Quadratic Unconstrained Binary Optimization (QUBO) and Ising Spin Glasses (Ising model). While the primary focus is on solving problems modeled as QUBOs, we will also provide a brief introduction to Ising models. Understanding the Ising model, the physics-based counterpart to QUBOs, is essential for appreciating their relevance in quantum computing. This connection highlights how problems in optimization can be directly mapped onto physical systems, enabling solutions through quantum techniques like quantum annealing.

\begin{figure}[htbp]
\centering
\resizebox{\linewidth}{!}{%
\begin{tikzpicture}[
    node distance=0.3cm and 1.0cm,
    every node/.style={font=\scriptsize, align=center},
    block/.style={rectangle, draw=darkred!80, fill=red!20, thick, text width=3cm, minimum height=0.8cm, rounded corners},
    smallblock/.style={rectangle, draw=darkviolet!80, fill=violet!20, thick, text width=3.5cm, minimum height=0.8cm, rounded corners},
    implblock/.style={rectangle, draw=darkblue!80, fill=blue!15, thick, text width=2cm, minimum height=0.8cm, rounded corners},
    arrow/.style={-Stealth, thick, darkblue}
]

% Level 1
\node[block] (qubo) {\hyperref[sec:qubo]{QUBO Model \\ (2)}};

% Level 2
\node[block, below left=of qubo] (np) {\hyperref[subsec:np]{Number Partitioning \\ (3.1)}};
\node[block, below=of qubo] (mc) {\hyperref[subsec:mc]{Max-Cut \\ (3.2)}};
\node[block, below right=of qubo] (mvc) {\hyperref[subsec:mvc]{Minimum Vertex Cover \\ (3.3)}};

% Level 3
\node[block, below=of np] (op) {\hyperref[subsec:mc]{Order Partitioning \\ (4.1)}};
\node[block, below=of mvc] (cg) {\hyperref[subsec:cg]{Cancer Genomics \\ (4.2)}};

\node[draw=orange, thick, dashed, rounded corners, fit=(qubo)(np)(mc)(mvc)(op)(cg), inner sep=0.2cm] (groupbox1) {};

% Level 4
\node[smallblock, below=of mc, yshift=-1.4cm] (found1) {\hyperref[subsec:found1]{Foundations I:\\ Building Blocks \\(5.1)}};
\node[smallblock, below=of found1, yshift=-0.2cm] (found2) {\hyperref[subsec:found2]{Foundations II:\\ Algorithms for QUBOs \\ (5.2)}};

% Level 5
\node[smallblock, below left=of found2, xshift=-0.2cm] (sa) {\hyperref[subsec:sa]{Simulated Annealing (Theory) \\ (5.2.1)}};
\node[smallblock, below=of found2] (qaoa) {\hyperref[subsec:qaoa]{Quantum Approximate \\ Optimization Algorithm \\ (Theory) \\ (5.2.3)}};
\node[smallblock, below right=of found2, xshift=0.2cm] (qa) {\hyperref[subsec:qa]{Quantum Annealing (Theory) \\ (5.2.2)}};

\node[draw=purple, thick, dashed, rounded corners, fit=(found1)(found2)(sa)(qa)(qaoa), inner sep=0.2cm] (groupbox) {};

% Level 6 (QAOA Implementations)
\node[implblock, below=of qaoa, xshift=-2.5cm, yshift=-0.35cm] (qaoa1) {\hyperref[subsec:vqaoa]{QAOA Custom\\(Implementation) (6.1)}};
\node[implblock, below=of qaoa, yshift=-0.35cm] (qaoa2) {\hyperref[subsec:oqaoa]{Open-QAOA\\(Implementation) (6.2)}};
\node[implblock, below=of qaoa, xshift=2.5cm, yshift=-0.35cm] (qaoa3) {\hyperref[subsec:qqaoa]{Qiskit-QAOA\\(Implementation) (6.3)}};
\node[implblock, below=of sa, yshift=-0.75cm] (sa2) {\hyperref[subsec:sai]{Simulated Annealing\\(Implementation) (6.4)}};
\node[implblock, below=of qa, yshift=-0.75cm] (qa2) {\hyperref[subsec:qai]{Quantum Annealing\\(Implementation) (6.5)}};

\node[draw=blue, thick, dashed, rounded corners, fit=(qaoa1)(qaoa2)(qaoa3)(sa2)(qa2), inner sep=0.1cm] (groupbox) {};

% Arrows
\draw[arrow] (groupbox1.south) -- (found1.north);
\draw[arrow] (qubo) -- (np);
\draw[arrow] (qubo) -- (mc);
\draw[arrow] (qubo) -- (mvc);

\draw[arrow] (np) -- (op);
\draw[arrow] (mvc) -- (cg);
\draw[arrow] (found1) -- (found2);

\draw[arrow] (found2) -- (sa);
\draw[arrow] (found2) -- (qa);
\draw[arrow] (found2) -- (qaoa);

\draw[arrow] (qaoa) -- (qaoa1);
\draw[arrow] (qaoa) -- (qaoa2);
\draw[arrow] (qaoa) -- (qaoa3);

\draw[arrow] (qa) -- (qa2);
\draw[arrow] (sa) -- (sa2);

\end{tikzpicture}%
}
\caption{Flow of topics.}
\label{fig:flow-of-topics}
\end{figure}

\section{Quadratic Unconstrained Binary Optimization (QUBO) Models}
\label{sec:qubo}
The Quadratic Unconstrained Binary Optimization (QUBO) model is a versatile framework that has been applied to various optimization problems in fields such as operations research, finance, physics, and machine learning. Although not all problems are naturally formulated as QUBO instances, the model provides a common structure that can represent a wide range of combinatorial optimization challenges through appropriate transformations. The mathematical matrix form of a QUBO is
\[
    \min_{\mathbf{x} \in \{0, 1\}^n} [ \mathbf{x}^{\text{T}}\mathbf{Q}\mathbf{x}+c],
\]
where $\mathbf{x} \in \{0, 1\}^{n}$ represents the solution vector, $\mathbf{Q} \in \mathbb{R}^{n \times n}$ is a symmetric matrix, and $c \in \mathbb{R}$ is a constant. (Note that $c$ does not make any difference to the optimal solution.)
After multiplying the matrix form, the resulting equation is a quadratic unconstrained binary optimization model (QUBO):
\[
    \min_{x_{i} \in \{0, 1\}}[\sum_{i,j}Q_{ij}x_{i}x_{j} + \sum_{i}Q_{ii}x^{2}_{i} + c].
\]
%Where $G$ is the graph representation and $V(G)$ and $E(G)$ are the corresponding vertex and edge set respectively. 
Since $\mathbf{x}$ is a binary vector, $x^2_{i}=x_{i}$, we obtain the following triangular form:
\[
        \min_{x_{i} \in \{0, 1\}}[\sum_{i < j}Q_{ij}x_{i}x_{j} + \sum_{i}Q_{ii}x_{i} + c].
\]
QUBO models were first introduced and systematically studied in the 1960s \citep{Hammer1968, Ginsburgh1969}. These models were initially popular due to their ability to represent various integer and combinatorial optimization problems, where some decision variables are subject to integer constraints. These combinatorial problems include NP-Hard problems which, by definition, have no known polynomial-time exact solving algorithm, so QUBO models provide researchers with an alternative and equivalent way to study these types of problem. 

Historically, a variety of classical algorithms have been employed to solve QUBO problems. These include branch-and-bound \citep{Pardalos1990}, semidefinite optimization \citep{Helmberg1998}, and interior point methods for $Q$ matrices that are positive semidefinite \citep{Abello2001}. More relevant to this work is the observation that several heuristic and metaheuristic algorithms---such as simulated annealing \citep{kirkp}, genetic algorithms \citep{10.1145/3194452.3194463}, and tabu search \citep{glover1989tabu}--have been effectively applied to QUBO problems. While these methods are designed for general-purpose optimization rather than targeting specific combinatorial problems, they have shown the potential to produce high-quality solutions across a range of QUBO instances. In some cases, they may even offer competitive performance compared to specialized algorithms, particularly in terms of solution quality or runtime.

\subsection{Quantum Landscape for QUBOs}

In recent years, the rise of quantum computing has sparked significant interest in Quadratic Unconstrained Binary Optimization (QUBO) models, driven by advances in gate-based and annealing-based quantum heuristics \citep{Gomes2024}. QUBOs have garnered attention because of their mathematical equivalence to physics-based Ising models, which are central to quantum annealing processes. The following sections provide a brief overview of four approaches to QUBO formulation discussed in the literature; however, our focus will be on the first two, owing to their generality, relative simplicity, and broader acceptance within the field.

\subsubsection{Gate-Based Quantum Computing}
Gate-based quantum computing has introduced novel approaches to solving QUBO problems. For example, Farhi et al. proposed the Quantum Approximate Optimization Algorithm (QAOA), which takes advantage of quantum gates to find approximate solutions to combinatorial optimization problems \citep{farhi2014quantum}, including those formulated as QUBOs. This approach has been explored for various optimization tasks, demonstrating the potential of gate-based quantum computing in addressing QUBO problems.

\subsubsection{Quantum Annealing}
Quantum annealing, as explored by Kadowaki and Nishimori, offers a method to find the global minimum of a function utilizing quantum fluctuations \citep{kadowaki1998quantum}. This approach is particularly suited for solving QUBO problems, as it can efficiently navigate the solution space to identify optimal or near-optimal solutions. D-Wave Systems has developed quantum annealers that have been applied to various optimization problems, showcasing the practical applicability of quantum annealing in solving QUBO models.

\subsubsection{Quantum-Classical Heuristics}

Recent studies have explored a variety of hybrid quantum-classical algorithms to tackle QUBO problems, aiming to exploit the complementary strengths of quantum subroutines and classical optimization. One notable example is the Quantum-Assisted Genetic Algorithm (QAGA) \citep{king2019quantum}, which uses quantum annealing as a mutation operator within a classical genetic algorithm. Other approaches include neural network-inspired variational methods such as Quantum Circuit Born Machines (QCBMs) and hybrid generative models \citep{benedetti2019generative}, which can encode QUBO landscapes into classically trained quantum circuits. Quantum kernel methods, such as those based on variational quantum feature maps, have also been applied to combinatorial optimization problems through support vector classification strategies \citep{schuld2019quantum}. In addition, reinforcement learning combined with quantum policy evaluation \citep{jerbi2023quantumrl} has been proposed to efficiently navigate large solution spaces. These emerging hybrid methods expand the design space for QUBO solvers beyond annealing or QAOA-based schemes, illustrating the growing versatility of quantum-classical co-design in optimization.

\subsubsection{Quantum-Inspired Algorithms}
In addition to quantum computing approaches, several quantum-inspired classical algorithms have been developed to solve QUBO problems. These algorithms, while not utilizing quantum hardware, incorporate principles from quantum mechanics to enhance optimization processes. Examples include Quantum Particle Swarm Optimization (QPSO) \citep{sun2004quantum}, Quantum Genetic Algorithms (QGA) \citep{narayanan1996quantum}, and Quantum-Inspired Evolutionary Algorithms (QEA) \citep{han2009quantum}, among others. These methods have been applied to various optimization tasks, demonstrating the versatility and effectiveness of quantum-inspired techniques in solving QUBO problems.

\section{Three Canonical Problems}
\label{sec:problems}
We have chosen three well-known canonical problems—\textit{Number Partitioning} \citep{Garey1979}, \textit{Max-Cut} \citep{Goemans1995}, and \textit{Minimum Vertex Cover} \citep{Zhou2022} -- to formulate as QUBO models and solve with quantum algorithms. These three problems are part of Karp's 21 NP-Complete Problems and thus are well known and studied in the field of theoretical computer science. The reader can refer to the literature that describes different techniques to compare with the approaches in this paper \citep{Karp1972, https://doi.org/10.48550/arxiv.2410.22810}. By definition, there are no known exact classical polynomial-time solutions to these problems, making them viable targets for quantum speedup.

\subsection{Number Partitioning}
\label{subsec:np}
%The Number Partitioning problem (NPP) is one of the six fundamental NP-Hard Problems identified by Gary and Johnson. It serves as the base for other NP-Hard Problems such as Knapsack and Bin Packing, and has direct applications in fields ranging from Load Balancing to Social Network Analysis \citep{article}. 

The Number Partitioning problem goes as follows: Given a set $S$ of positive integer values $\{s_1, s_2, s_3 ... s_n\}$, partition $S$ into two sets $A$ and $S \setminus A$ such that
$$d=|\sum_{s_i \in A}s_{i}-\sum_{s_j \in S \setminus A}s_{j}|$$
is minimized. In simple words, the goal is to partition the set into two subsets where the sums are as close in value to each other as possible.

This objective function can be expressed as a QUBO using the following binary variable: $x_i=1$ indicates that $s_i$ belongs to $A$, and if $x_{i} = 0$ then $s_i$ in $S \setminus A$. The sum of elements in $A$ is $\sum_{i=1}^{n}s_ix_i$ and the sum of elements in $S \setminus A$ is $c-\sum_{i=1}^{n}s_ix_i$, where $c$ is the sum of elements in $S$, a constant. Thus, the difference in the sums is:
$$d=c - 2\sum_{i=1}^{n}s_ix_i.$$
This difference is minimized by minimizing the QUBO:
$$d^2=(c-2\sum_{i=1}^{n}s_ix_i)^2=c^2+4\mathbf{x}^{\text{T}}\mathbf{Q}\mathbf{x}$$
with $q_{ij}$ in $\mathbf{Q}$ defined below:
$$q_{ij}= \begin{cases}
  s_{i}(s_{i}-c) & \text{if } i = j \\
  s_{i}s_{j} & \text{if } i \neq j.
\end{cases}
$$

\subsection{Max-Cut}
\label{subsec:mc}
%The Max-Cut Problem is another famous NP-Hard problem that has been extensively studied with both classical and quantum algorithms. There are a variety of different techniques with different guaranteed approximation ratios ranging from Semidefinite Programming with Goemans-Williamson \citep{article1} to Metaheuristics like Tabu Search \citep{glover1989tabu}. Even the original Quantum Approximate Optimization Algorithm was used to solve the Max-Cut Problem on 2-regular and 3-regular graphs \citep{farhi2014quantum}. Max-Cut also has countless applications across various domains ranging from machine learning to theoretical physics.

The Max-Cut problem is the following: Given an undirected graph, $G$ with a vertex set $V$ and an edge set $E$, partition $V$ into sets $A$ and $V \setminus A$ such that the number of edges connecting nodes between these two sets is maximized. See Figure \ref{fig:maxcut-inst} for a graph with six nodes labeled $0,1,2,3,4,5$ and eight edges.

Similarly to Number Partitioning, the QUBO can be constructed by setting $x_i=1$ if the vertex $i$ is in $A$ and letting $x_i=0$ if the vertex $i$ is in $V \setminus A$. The expression $x_{i}+x_{j}-2x_{i}x_{j}$ identifies whether the edge $(i, j) \in E$ is in the cut:  $x_{i}+x_{j}-2x_{i}x_{j}$ is  $1$ if and only if exactly one of $x_{i}$ or $x_{j}$ is $1$ (and the other is $0$). The objective function is:
\[\max[\sum_{(i,j) \in E}x_{i}+x_{j}-2x_{i}x_{j}].\]
Since QUBOs are typically framed as minimization problems, the Max-Cut QUBO is:
\[\min[\sum_{(i,j) \in E}2x_{i}x_{j}-(x_{i}+x_{j})].\]
Note that this can be expressed as
\[\min \textbf{y} = \textbf{x}^{\text{T}}\mathbf{Q}\textbf{x}\]
with a matrix $\textbf{Q}$ where the linear terms represent the diagonal and the quadratic terms represent the off-diagonal elements. 
% The Maximum Cut Problem is one of the quintessential combinatorial optimization problems \citep{Commander2009} that has been studied in a variety of contexts, notably being the target of different approximation algorithms such as the famous Goemans-Williamson algorithm \citep{Goemans1995} that inspired the semidefinite programming paradigm \citep{Gärtner2012} and the original QAOA algorithm \citep{farhi2014quantum}.

\subsection{Minimum Vertex Cover}
\label{subsec:mvc}
%The Minimum Vertex Cover (MVC) is the last well-studied NP-Hard problem that we will discuss in this paper. Similar to the Number Partitioning Problem, the decision version of Minimum Vertex Cover was one of Karp's 21 NP-Complete Problems, and like Max-Cut it has some well-known approximation algorithms. The Minimum Vertex Cover problem also has many applications in fields ranging from cybersecurity to metabolic engineering \citep{Pelofske_2019}.

As the name implies, the Minimum Vertex Cover problem seeks to find the minimum vertex cover of an undirected graph $G$ with vertex set $V$ and edge set $E$. A vertex cover is a subset of vertices such that each edge $(i, j) \in E$ shares an endpoint with at least one vertex in the subset.

The Minimum Vertex Cover problem has two features:
\begin{enumerate}
    \item the number of vertices in the vertex cover must be minimized, and
    \item the edges must be covered by the vertices. (All edges must share at least one vertex from the subset of vertices.)
\end{enumerate}

If the inclusion of each vertex in the vertex cover is denoted with a binary value ($1$ if it is in the cover, and $0$ if it is not), the minimal vertex property can be expressed as the term
\[Q_{1} = \sum_{i \in V} x_{i}.\]
Similar to  Max-Cut, the covering criterion is expressed with constraint (for all $(i,j) \in E$):
\[x_{i} + x_{j} \geq 1.\]
This can be represented with the term and the use of a penalty factor $P$:
\[Q_{2} = P \cdot (\sum_{(i, j) \in E}(1-x_{i}-x_{j}+x_{i}x_{j})).\]
Thus, the QUBO for Minimum Vertex Cover is to minimize:
\[\sum_{i \in V} x_{i} + P \cdot (\sum_{(i, j) \in E}(1-x_{i}-x_{j}+x_{i}x_{j})).\]
Unlike the other formulations, our model incorporates a penalty factor \( P \) that enforces the covering criteria. It is crucial to select \( P \) appropriately; if \( P \) is too small, the formulation may not accurately represent the problem, leading to solutions that do not meet the requirements of the application. Conversely, if \( P \) is too large, the model can become ill-conditioned, resulting in numerical instability and potentially violating hardware limitations in the coefficient ranges. Therefore, choosing an appropriate \( P \) is essential to ensure that the formulation correctly models the problem and is solvable on quantum hardware. Optimal parameter selection is a challenging problem and the parameters chosen in this tutorial were done so in a heuristic fashion; however, interested readers can refer to the following papers for more detail \citep{vyskocil2022optimal, karimi2017boosting, Ayodele2022, Glover2018, lucas2014ising}.
\section{Two Practical Problems}
\label{sec:practproblems}
We now introduce two practical problems.
%that are novel. As a result, there is no pre-existing literature on these problem formulations and their applications.
The Ordering Partitioning problem is described for the first time in this tutorial. It is different, but 
%; thus, no reference materials discuss formulations or techniques. 
related to the well-studied Portfolio Optimization problem \citep{10087257, Sodhi22} and is a natural extension of the Number Partitioning problem, so it offers a graceful transition from canonical problems to practical ones. The Cancer Genomics problem identifies altered cancer pathways using mutation data from the Cancer Genome Atlas (TCGA) acute myeloid leukemia (AML) data set and is related to a different canonical problem, the Maximum Independent Set problem, which reduces to the Vertex Cover problem \citep{Alghassi845719, Ley2013GenomicAE}.

\subsection{Order Partitioning}
\label{subsec:op}
%This final application and problem discussed is a variation of the pre-existing portfolio optimization problem named Order Partitioning.
%The objective of the Order Partitioning Problem is to partition a set of stocks into two groups where the difference between the net prices of the stocks and total risks are minimized. 
Order\footnote{We retain the industry term \textit{order} for a block of stock that is traded. In our context of A/B testing, these blocks cannot be split.} Partitioning is used for A/B testing of various investment strategies (proposed by hedge fund researchers). Also known as split testing, A/B makes variable-by-variable modifications and allows firms to optimize their strategies while avoiding the risks of testing at a larger scale. Order Partitioning  reduces to Number Partitioning and thus is an NP-complete. This class of problems by definition has no good polynomial-time algorithms and is a candidate for quantum speedup \citep{Williams2011}.

The Order Partitioning problem goes as follows: We are given a set of $n$ stocks each with an amount $q_j$ dollars totaling $T$ dollars ($\sum_{j=1}^{n}q_j=T$). There are $m$ risk factors. Let $p_{ij}$ be the risk exposure to factor $i$ for the stock $j$. The objective is to divide stocks $j=1 \ldots n$ into sets $A$ and $B$ such that:

\begin{enumerate}
    \item the sizes are equal ($\sum_{j \in A} q_j = \sum_{i \in B} q_j = \frac{T}{2}$) and
    \item the risks are equal (for each factor $i= 1 \ldots m$: 
    $\sum_{j\in A}p_{ij} = \sum_{j\in B}p_{ij}$).
\end{enumerate}

If exact equality is not possible, then we want the absolute difference to be minimized:
\begin{enumerate}
    \item $\min |\sum_{j \in A} q_j - \sum_{j \in B} q_j|$ and
    \item for each factor $i = 1 \ldots m$: 
    $\min|\sum_{j\in A}p_{ij} -\sum_{j\in B}p_{ij}|.$
\end{enumerate}

Since Objective 1 for each case is the same as the Number Partitioning problem, we can reuse the QUBO derived earlier (replacing $s$ with $q$, and noting that $x_j$ is binary):
\[Q_{1}=(T-2\sum_{j=1}^{n}q_{j}x_{j})^2.\]
To satisfy Objective 2 for each case, we would first like to reduce the absolute risk on each stock before minimizing its overall stocks. Consider the sign variables $\sigma_{i}$ that only take values in $\{-1, 1\}$; the minimum absolute risk is $0$. This case is similar to the Number Partitioning problem where absolute risk is minimized by minimizing the square of the sum:
\[\min (\sum_{j=1}^{n}p_{ij}\sigma_{j})^2.\]
Considering the minimum risk in all stocks, we get the following final expression to minimize the risk factors using the terms of the sign variables:
\[\min \sum_{i=1}^{m}(\sum_{j=1}^{n}p_{ij}\sigma_{j})^2.\]
% To partition each of the risk factors we can briefly consider Ising spins to minimize the magnitude of the following matrix $\Vec{h}$. Note that as mentioned earlier $p_{ij}$ is the risk factor $i$ for stock $j$ and $\sigma_{j}$ is the Ising Spin corresponding to stock j.
% \[
%     \Vec{h} = 
% \begin{pmatrix}
% p_{11} & p_{12} & p_{13} & \cdots & p_{1n} \\
% p_{21} & p_{22} & p_{23} & \cdots & p_{2n} \\
% p_{31} & p_{32} & p_{33} & \cdots & p_{3n} \\
% \vdots & \vdots & \vdots & \vdots & \vdots \\
% p_{m1} & p_{m2} & p_{m3} & \cdots & p_{mn}
% \end{pmatrix}
% \begin{pmatrix}
% \sigma_{1} \\
% \sigma_{2} \\
% \sigma_{3} \\
% \vdots \\
% \sigma_{n}
% \end{pmatrix}
% \]

\noindent Converting the sign variables to the QUBO variables $x_{i} $ (through transformation $\sigma_{i} = 2x - 1$) and weighting with the penalty factor $P$, we get
\[Q_{2} = P\sum_{i=1}^{m}(\sum_{j=1}^{n}p_{ij}(2x_{j}-1))^2.\]

\noindent The complete QUBO for Order Partitioning is
\[Q = (T-2\sum_{j=1}^{n}q_{j}x_{j})^2 + P\sum_{i=1}^{m}(\sum_{j=1}^{n}p_{ij}(2x_{j}-1))^2.\]

\noindent Note that $P$ depends on how strictly each constraint is to be enforced. %For a relaxed implementation of reducing risk factors, $B$ is decreased, and the opposite holds as well.

\subsection{Cancer Genomics using TCGA}
\label{subsec:cg}
%\subsubsection*{Identifying Single Cancer Pathway}
The second application is the de novo identification of altered cancer pathways from shared mutations and gene exclusivity. A cancer pathway is an identified sequence of genes whose identification is useful for understanding the disease and developing treatments. This problem non-trivially reduces to the Independent Set (which is well known to reduce from the Vertex Cover \citep{10.1007/978-3-319-42634-1_28}). Thus, this practical problem is also NP-complete and is a suitable candidate for quantum speedup. \\
Patients and their corresponding gene mutations are first modeled as the incidence matrix $B$ of a hypergraph with each vertex $g_{i}$ representing a gene and each patient $P_{i}$ representing the hyperedge. Note that this incidence matrix is of the form
\[B = \begin{pmatrix}
b_{11} & b_{12} \ldots b_{1m} \\
b_{21} & b_{22} \ldots b_{2m} \\
\vdots & \vdots \ddots \vdots \\
b_{n1} & b_{n2} \ldots b_{nm}
\end{pmatrix}\]
where each of the $m$ columns represents each patient $P_{i}$ mutation gene list, and each of the $n$ rows represents the presence of gene $g_{j}$ in a patient's mutated gene list. For example, if gene $g_{i}$ is mutated for patient $P_{j}$, then $b_{ij} = 1$ otherwise $b_{ij} = 0$. Using this incidence matrix, we can derive the graph Laplacian as follows:
\[L^{+} = BB^{\text{T}}.\]
\noindent Two key combinatorial criteria are essential to identifying driver mutations:
\begin{itemize}
    \item Coverage: We want to identify which genes are the most prevalent among cancer patients. If a gene is shared between many cancer patients, it is more likely that the gene is a driver gene.
    \item Exclusivity: If there is already an identified cancer gene on the patient's gene list, it is less likely there will be another. This is not a hard rule and is sometimes violated.
\end{itemize}
Based on the criteria above, we obtain the following two matrices, decomposing the Laplacian graph $L$ as
\[L^{+} = \mathbf{D} + \mathbf{A}.\]
\begin{itemize}
    \item Degree Matrix $\mathbf{D}$: This diagonal matrix corresponds to the coverage criterion. Each index $d_{ij}$, where $i=j$ represents the number of patients affected by the gene $i$. All other entries are $0$, and this attribute should be maximized.
    \item Adjacency matrix $\mathbf{A}$: This adjacency matrix corresponds to the exclusivity criterion. Each $a_{ij}$ with $i \neq j$ represents the number of patients affected by the gene $i$ and the gene $j$. All other entries are $0$, and this attribute should be minimized.
\end{itemize}
Let us begin by trying to find just one pathway. We define a solution pathway as $\mathbf{x} = \begin{pmatrix}
    x_{1} & x_{2} & x_{3} & ... & x_{n}
\end{pmatrix}^{\text{T}}$
where for all $i \in \{1, 2, 3, \dots, n\}$, $x_{i}$ is a binary variable. If $x_{i}=0$, then the gene $i$ is not present in the cancer pathway. If $x_{i}=1$, the gene is present in the cancer pathway. The exclusivity term is $\mathbf{x}^{\text{T}}\mathbf{A}\mathbf{x},$ and the coverage term is $\mathbf{x}^{\text{T}}\mathbf{D}\mathbf{x}$. Since we want the exclusivity term to be minimized and the coverage term to be maximized, the QUBO to identify cancer pathways from a set of genes is
\[\mathbf{x}^{\text{T}}\mathbf{A}\mathbf{x} - \alpha \mathbf{x}^{\text{T}}\mathbf{D}\mathbf{x},\]
where $\alpha$ is a penalty coefficient. Since coverage is more important than exclusivity, we weigh it more, and $\alpha \geq 1$. This QUBO can be written equivalently using the values $a_{ij}$ and $d_{ij}$ of $A$ and $D$, respectively, as follows:
\[\sum_{i=1}^{n}\sum_{j=1}^{n}a_{ij}x_{i}x_{j}-\alpha\sum_{i=1}^{n}d_{i}x_{i}.\]
%\subsubsection{Identifying Multiple Cancer Pathways}
This QUBO can be extended to identify multiple cancer pathways in a single run \citep{article2}. Note that this is just one possible formulation to formulate the cancer pathway problem; however, other approaches exist. See, for example, \citep{10.1007/978-3-642-20036-6_44, Zhao2012EfficientMF}.
%  Let $\mathbf{X} = \begin{pmatrix}
%     \mathbf{x_{1}} & \mathbf{x_{2}} & ... &
%     \mathbf{x_{k}}
% \end{pmatrix}^{\text{T}}$
% where each $i \in \{1, 2 ... k\}$, $\mathbf{x}_{i}$ represents a cancer pathway. Define $\mathbf{L}=\mathbf{A}+\mathbf{D}$, denote $\mathbf{I_{k}}$ as the $k \times k$ identity matrix, denote $\mathbf{I_{n}}$ as the $n \times n$ identity matrix, and denote $\mathbf{J_{k}}$ as the $k \times k$ matrix of $1$s. With the use of tensor products $\mathbf{Q}_{main} = - \mathbf{I_{k}} \otimes \mathbf{L}$ and $\mathbf{Q}_{orth} = (\mathbf{J_{k}} - \mathbf{I_{k}}) \otimes \mathbf{I_{n}}$, the final QUBO is:
% \[\mathbf{X}^{\text{T}}(\mathbf{Q}_{main} + \alpha \mathbf{Q}_{orth})\mathbf{X}.\]

%Similar to the single pathway case $\alpha$ represents the penalty factor.

\section{Moving from Classical Optimization to Quantum Computing}
\label{sec:class-to-quantum}

%Readers are advised to use this section to get a fundamental understanding of how quantum computing works, however, it is not essential to know every little detail. In the same way, programmers don't need to understand how every bit is manipulated when writing an algorithm, quantum programmers don't need to visualize all qubit operations. That being said it will be useful to read this section or similar materials to get a better understanding of the subject matter. Also, quantum computing is an expansive and growing field and there exist many more interesting nuanced concepts beyond the ones discussed here, however, this context is sufficient to understand the rest of the paper.

This primer offers a concise introduction to quantum computing, exploring its two primary paradigms: gate-based quantum computing and adiabatic quantum computing. Although both paradigms are theoretically equivalent in computational power, they differ fundamentally in their approaches. We will focus on the core algorithms of each model, investigating and implementing them in detail. The gate-based model utilizes the Quantum Approximate Optimization Algorithm (QAOA) to tackle combinatorial optimization problems, whereas the adiabatic quantum computing model is realized primarily through quantum annealing.

This section is divided into two parts. The first part covers essential concepts and principles in quantum mechanics and quantum computing, providing the foundation needed to understand the relevant algorithms. The second part delves into algorithms designed to solve Quadratic Unconstrained Binary Optimization (QUBO) problems. We begin the second subsection with a discussion of simulated annealing, a well-known classical algorithm in the Operations Research community, before introducing quantum annealing and QAOA, along with their key concepts and applications.

\subsection{Foundations of Quantum Computing (Part I): Building Blocks}
\label{subsec:found1}

Part I introduces fundamental concepts in quantum computing, focusing on quantum bits (qubits) and their properties. It explains spins as the intrinsic angular momentum of quantum particles and how superposition allows quantum systems to exist in multiple states simultaneously. The section also covers the behavior of qubits, which can represent a combination of states, unlike classical bits. Additionally, it touches on the concept of quantum gates, which manipulate qubits through unitary operators, enabling quantum circuits. These foundational building blocks provide the necessary understanding for further exploration of quantum algorithms and computational models.
\subsubsection{Spins}
\label{subsubsec:spins}
Quantum spin is a fundamental property of quantum particles such as electrons, protons, and neutrons that reflect intrinsic angular momentum. Unlike classical angular momentum, quantum spin does not correspond to actual rotation, but governs how a particle interacts with magnetic fields and other particles. Spin values are quantized and are typically described as ``spin-up" with a spin value of \(+\frac{1}{2}\) or "spin-down" with a spin value of \(-\frac{1}{2}\).

\subsubsection{Superposition}
\label{subsubsec:superposition}
Unlike classical systems, which must be in a single deterministic state, quantum systems can exist in a superposition: a combination of multiple states simultaneously. For example, the quantum spin of a particle \(p\) can be described as
\[
p = \alpha \ket{\uparrow} + \beta \ket{\downarrow},
\]
where \(\uparrow\) represents spin-up, \(\downarrow\) represents spin-down, and \(\alpha\) and \(\beta\) are the probability amplitudes for measuring each of these spin states. While the particle in superposition is described as having both spin-up and spin-down, once measured, the superposition collapses. The spin will collapse into spin-up with a probability of \(|\alpha|^2\), or into spin-down with a probability of \(|\beta|^2\).

Note that in this section and the following text, vectors are expressed as ``kets": $\ket{\psi}$. Their adjoints are known as ``bras": $\bra{\psi} = \overline{\ket{\psi}}^{\text{T}}$ and are also frequently used in calculations. This notation is known as the Dirac notation and is commonly used in quantum mechanics.

\subsubsection{Qubits}
\label{subsubsec:qubits}
In quantum computing, the fundamental unit of information is the quantum bit, or qubit. Qubits have two basic basis states: \(\ket{0}\), typically implemented with a quantum spin-up, and \(\ket{1}\), typically implemented with a quantum spin-down. Classical bits are restricted to the values 0 or 1. Qubits, leveraging quantum spins, can exist in a superposition of both states. A qubit's state is described as:
\[
\ket{\psi} = \alpha \ket{0} + \beta \ket{1},
\]
where \(\alpha\) and \(\beta\) are complex coefficients that represent the probability amplitudes for the states \(\ket{0}\) and \(\ket{1}\), respectively. This superposition allows qubits to represent both 0 and 1 simultaneously, enabling quantum computers to process information in parallel and solve certain problems more efficiently than classical computers.

Mathematically, $\ket{0}$ and $\ket{1}$ are basis states that are vectors in a 2-dimensional Hilbert Space:\[\ket{0}=\begin{pmatrix}
    1 \\
    0 \\
\end{pmatrix} \hspace{0.5cm} \ket{1}=\begin{pmatrix}
    0 \\
    1 \\
\end{pmatrix}.\]
The underlying mathematics goes beyond the scope of the paper, but a Hilbert Space can be understood as a special type of (possibly infinite dimensional) vector space\footnote{See an earlier tutorial by \cite{SiddhuTayur} for details.}.
\begin{figure}[h]
  \centering
\includegraphics[width=0.5\textwidth]{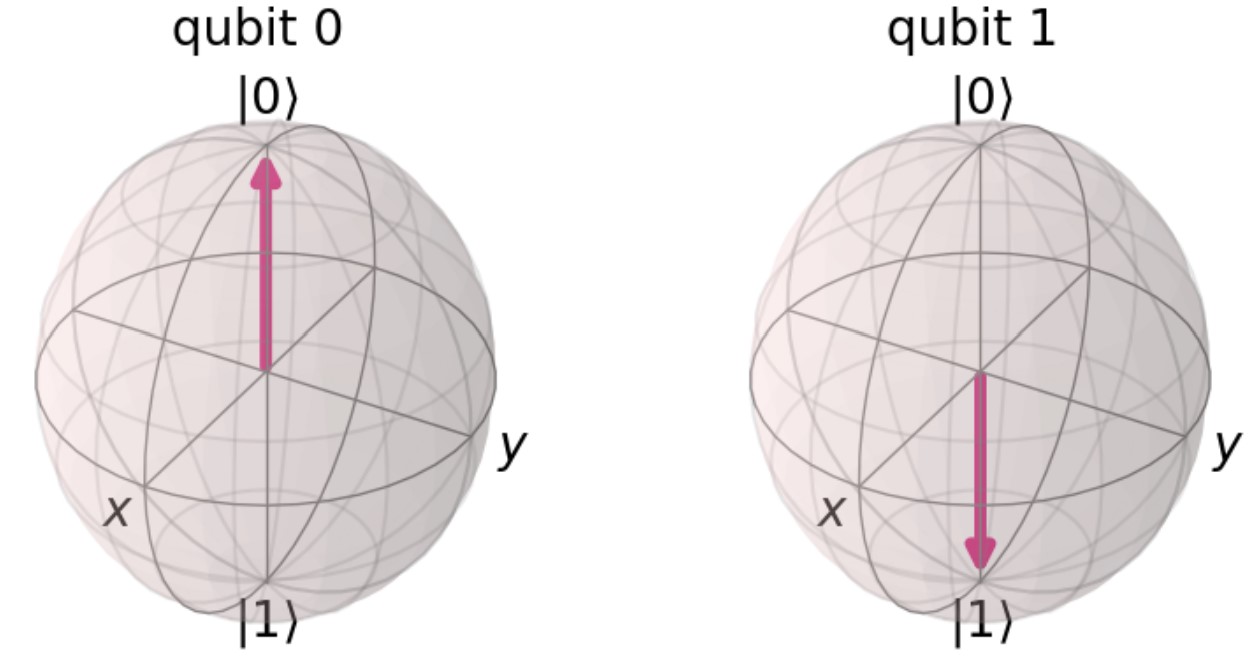}
  \caption{Bloch Sphere representations of $\ket{0}$ and $\ket{1},$ respectively.}
  \label{fig:block_sphere}
\end{figure}
In the Bloch sphere in Figure  \ref{fig:block_sphere}, the $x$ and $y$ axes represent the real and imaginary parts of the probability amplitude, respectively. The $z$-axis represents the probability difference between the $\ket{0}$ and $\ket{1}$ states.

The coefficients $\alpha$ and $\beta$ are complex-valued probability amplitudes that, when squared, represent the probability of measuring each state. This concept is known as Born's Rule and can be used to determine the probability of measuring each basis state:
$$P(0)=|\alpha|^2, P(1)=|\beta|^2,$$
$$|\alpha|^2+|\beta|^2=1.$$
Because of this ability for a qubit to probabilistically represent two states at once, a quantum register of $n$ qubits could have a superposition of $2^n$ states at once, with each state occurring with a certain probability. Below is the uniform distribution of $n$ qubits, where $\ket{i}$ represents the $i$th bitstring:
\[
    \frac{1}{\sqrt{2^n}}\sum^{2^n-1}_{i=0}\ket{i}.
\]
A classical register, on the other hand, would only be able to express $n$ states. This ability of qubits to map multiple states through superposition contributes to quantum speedup.

When multiple quantum states (each in different Hilbert spaces $\mathcal{H}_{i}$) or registers are combined, such as $\ket{\psi_1} \in \mathcal{H}_{1}$, $\ket{\psi_2} \in \mathcal{H}_{2}$,.., $\ket{\psi_n} \in \mathcal{H}_{n}$, the resulting system is the tensor or Kronecker product of the states: $\ket{\psi_1} \otimes \ket{\psi_2} \otimes ... \otimes \ket{\psi_n}=\ket{\psi_1\psi_2...\psi_n} \in \mathcal{H}_{1} \otimes \mathcal{H}_{2} \otimes \mathcal{H}_{3} \otimes\ldots\otimes\mathcal{H}_{n}$. Below is an example of taking the tensor product of a $\ket{0}$ and $\ket{1}$:
\[
    \ket{01}=\ket{0} \otimes \ket{1} = \begin{pmatrix}
    1 \\
    0 \\
\end{pmatrix} \otimes \begin{pmatrix}
    0 \\
    1 \\
\end{pmatrix} =
\begin{pmatrix}
    1 \cdot \begin{pmatrix}
    0 \\
    1 \\
\end{pmatrix} \\
    0 \cdot \begin{pmatrix}
    0 \\
    1 \\
\end{pmatrix} \\
\end{pmatrix} =
\begin{pmatrix}
    0 \\
    1 \\
    0 \\
    0
\end{pmatrix}.\]

\subsubsection{Quantum Gates}
\label{subsubsec:gates}
Qubits are manipulated using unitary operators (or matrices) $U$. These operators are defined by two key properties:
\begin{enumerate}
    \item They  map unit vectors to unit vectors.

    \item They are normal with $U^{\dagger}=U^{-1}$, so $UU^{\dagger}=U^{\dagger}U=I.$
\end{enumerate}

These unitary operators are known as quantum gates. Note that these operators can be visualized as reflections and rotations of qubit vectors, and, unlike classical computing, are all reversible. Below is a brief list of single qubit quantum gates and functionality.

\noindent The trivial Identity gate $I$ is often ignored when representing and implementing circuits, but it is necessary for mathematics. It preserves the state and is represented by the identity matrix:
$$I = \begin{pmatrix}
        1 & 0 \\
        0 & 1
    \end{pmatrix}.$$

\noindent The Hadamard gate $H$ is used to convert the basis states to a uniform superposition. This makes it equally likely to be measured in the $\ket{0}$ or $\ket{1}$ states, The matrix is shown below:
\[H = \frac{1}{\sqrt{2}}\begin{pmatrix}
        1 & 1 \\
        1 & -1
    \end{pmatrix}.\]
\noindent The Pauli gates $\sigma_{X}$ (or $X$), $\sigma_{Y}$ (or $Y$), and $\sigma_{Z}$ (or $Z$) are essential gates to rotate the quantum state vectors $180$ across their corresponding axes on the Bloch sphere in Figure \ref{fig:block_sphere}.
The three matrices are shown below:
\[\sigma_{X} = \begin{pmatrix}
        0 & 1 \\
        1 & 0
    \end{pmatrix}, \hspace{0.4cm}
\sigma_{Y} = \begin{pmatrix}
        0 & -i \\
        i & 0
    \end{pmatrix}, \hspace{0.4cm}
\sigma_{Z} = \begin{pmatrix}
        1 & 0 \\
        0 & -1
    \end{pmatrix}.\]
An important gate that arbitrarily rotates qubits by a phase shift $\theta$ without flipping bits is the phase gate $P$. This gate can also be understood to be a generalization of the $Z$ gate and has the following matrix representation:
$$ P(\theta) = \begin{pmatrix}
    1 & 0 \\
    0 & e^{i \theta}
\end{pmatrix}.$$
\noindent The Rotation Gates $R_X$, $R_Y$, and $R_Z$ are similar to the Pauli Gates but implement parametrized rotations over their respective axes as shown in their matrices below:
$$R_{X}(\theta) = \begin{pmatrix}
        \cos(\frac{\theta}{2}) & -i\sin(\frac{\theta}{2}) \\
        -i\sin(\frac{\theta}{2}) & \cos(\frac{\theta}{2})
    \end{pmatrix}, \hspace{0.2cm}
R_{Y}(\theta) = \begin{pmatrix}
        \cos(\frac{\theta}{2}) & -\sin(\frac{\theta}{2}) \\
        \sin(\frac{\theta}{2}) & \cos(\frac{\theta}{2})
    \end{pmatrix}, \hspace{0.2cm}
R_{Z}(\theta) = \begin{pmatrix}
        1 & 0 \\
        0 & e^{i\theta}
    \end{pmatrix}.$$
There also exists multi-qubit gates. The $CX$ (also known as $CNOT$) and $CZ$ read the input from a control qubit and then apply the corresponding $X$ or $Z$ operation. Their matrices are shown below:
$$
 CX =
\begin{pmatrix}
1 & 0 & 0 & 0 \\
0 & 1 & 0 & 0 \\
0 & 0 & 0 & 1 \\
0 & 0 & 1 & 0 \\
\end{pmatrix}, \hspace{0.5cm}
CZ =
\begin{pmatrix}
1 & 0 & 0 & 0 \\
0 & 1 & 0 & 0 \\
0 & 0 & 1 & 0 \\
0 & 0 & 0 & -1 \\
\end{pmatrix}.
$$
Another relevant multi-qubit gate for QAOA is the $U_{ZZ}(\theta)$ gate. This gate applies a phase to qubits based on the strength of their correlations and is useful for modeling costs between the two-body interactions present in QUBO models. Mathematically, it applies a phase $\theta$ (rotates it by $\theta$) when both qubits are $1$ (in the $Z \otimes Z$ state). Thus, it is given by the equation $U_{ZZ}(\theta) = e^{i \theta (Z \otimes Z)}$ or the following matrix representation:
$$
U_{ZZ}(\theta) =
\begin{pmatrix}
e^{i\theta} & 0 & 0 & 0 \\
0 & e^{-i\theta} & 0 & 0 \\
0 & 0 & e^{-i\theta} & 0 \\
0 & 0 & 0 & e^{i\theta} \\
\end{pmatrix}.
$$
There exist other single and multi-qubit quantum gates, but these are the only ones essential to this tutorial.

Theoretically, it has been proven that there exist universal gate sets $S$ (i.e., that any unitary transformation on an arbitrary number of qubits can approximate finite precision using only finite sequence gates from $S$).  One such universal gate set is $\{R_{X}(\theta), R_{Y}(\theta), R_{Z}(\theta), CX, P(\theta)\}$ \citep{Williams2011}. Universal gate sets are an open field of study that can be further explored in the following references \citep{Sawicki2022, aharonov2003simpleprooftoffolihadamard}.
\subsubsection{Quantum Circuits}
\label{subsubsec:circuits}
Similarly to combining a system of qubits, multiple operators can be combined over multiple qubits using the tensor product and represented as a quantum circuit. We illustrate with a simple example.

\noindent Let $q[0] = \ket{0}$ and $q[1]=\ket{0}$. Consider the operator formed by the expression:
\[[(CX) \times (I \otimes H)]\ket{00}.\]

\begin{figure}[h]
  \centering
\includegraphics[width=0.3\textwidth]{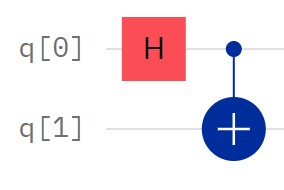}
  \caption{$H$ is Hadamard gate. The multi-qubit gate represented by the circle is $CX$ gate. }
  \label{fig:bell-state}
\end{figure}

%For example, for the simple Bell state circuits shown below from IBM Composer:
This is represented as a circuit shown in Figure~\ref{fig:bell-state}. Note that although the circuits are read from left to right, it is written as a matrix expression from right to left because of matrix notation. This order is important because of the noncommutativity of matrix multiplication. 

\begin{figure}[h]
  \centering
\includegraphics[width=0.4\textwidth]{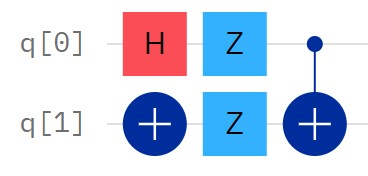}
  \caption{In addition to the gates in Figure \ref{fig:bell-state}, here we use $Z$ to represent a Pauli gate.}
  \label{fig:fourth-bell-state}
\end{figure}

\noindent As a second example, consider the expression (where $q[0] = \ket{0}$ and $q[1]=\ket{0}$):
\[[(CX) \times (Z \otimes Z) \times (X \otimes H)]\ket{00}.\]
This is represented by the circuit shown in Figure~\ref{fig:fourth-bell-state}.

% \subsubsection{Entanglement}

% %Note that two states $\ket{\phi^{+}}$ and $\ket{\Psi^{-}}$ are interesting due to being Bell states. Bell states are the simplest example of entangled states.
% Mathematically, entanglement means that a state cannot be expressed as the tensor product of single qubit states. For example, in an arbitrary entangled state $\ket{\psi} \in \mathcal{H}_{1} \otimes \mathcal{H}_{2}$, we cannot write
% \[\ket{\psi} = \ket{\psi_{1}} \otimes \ket{\psi_{2}}\]
% with $\ket{\psi_{1}} \in \mathcal{H}_{1}$ and $\ket{\psi_{2}} \in \mathcal{H}_{2}$.

% Intuitively, quantum entanglement happens when multiple qubits are correlated such that their states are dependent on each other. For example, with what is called the the First Bell state, $\ket{\phi^{+}}=\frac{1}{\sqrt{2}}(\ket{00}+\ket{11})$, if the first qubit is measured as $0$, the second one would also come up as $0$. Similarly, if the first qubit is $1$, the second would also be $1$. This property has been experimentally verified and offers potential for quantum speed-up. The circuit modelling this entangled state is represented in Figure~\ref{fig:bell-state}.

\subsubsection{Hamiltonian}
\label{subsubsec:hamiltonian}
The Hamiltonian \( H \) is a quantum-mechanical operator represented by a matrix that describes the total energy of a system. It has several key properties:
\begin{itemize}
    \item It is Hermitian (\( H^{\dagger} = H \)), ensuring that its eigenvalues are real.
    \item Its eigenvalues represent the possible energy levels of the system, while the corresponding eigenvectors describe the quantum states.
    \item It governs the time evolution of the quantum state of the system \( \ket{\psi(t)} \) via the Schrödinger equation, which is essential in quantum annealing: \[
    i \hbar \frac{\partial}{\partial t} \ket{\psi(t)} = H \ket{\psi(t)}.
    \]
\end{itemize}
\subsubsection{Ising Model}
\label{subsubsec:ising}
Ising models are mathematical tools originally introduced by Ernst Ising and Wilhelm Lenz to model ferromagnetism in statistical physics. The Ising model Hamiltonian is given by:
\[
    H(\sigma) =  -\sum_{i,j} J_{ij} \sigma_i \sigma_j - \sum_i h_i \sigma_i.
\]  

\begin{figure}[h!]
    \centering
    \begin{subfigure}[b]{0.45\textwidth}
        \centering
        \includegraphics[width=\linewidth]{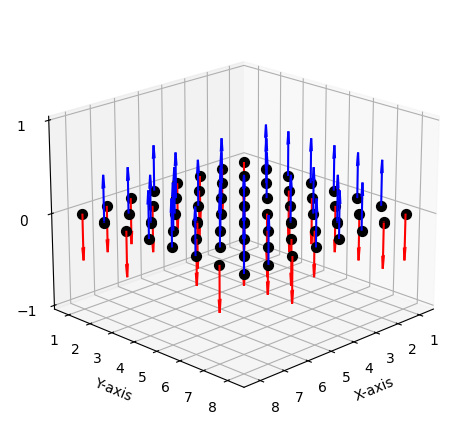}
        \caption{Ising Model of a Ferromagnet}
        \label{fig:image1}
    \end{subfigure}
    \hfill
    \begin{subfigure}[b]{0.35\textwidth}
        \centering
        \includegraphics[width=1\linewidth]{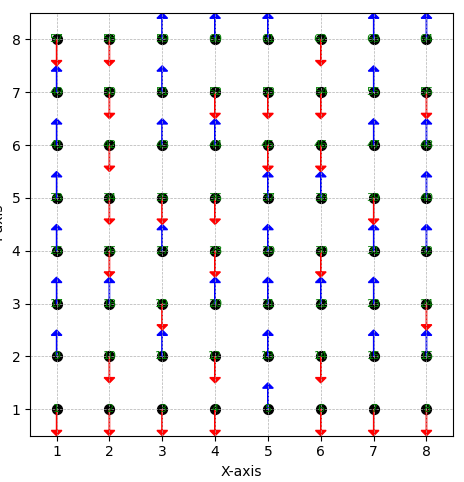}
        \caption{Top view of Ising Model}
        \label{fig:image2}
    \end{subfigure}
    \caption{Two-Dimensional Square-Lattice Ising model.}
    \label{fig:lattice}
\end{figure}

The system modeled by the Ising formulation is typically represented as a lattice graph. An example is shown in Figure \ref{fig:lattice}. Each spin variable \( \sigma_i \in \{-1, 1\} \) (different from \ref{subsubsec:spins}) corresponds to the quantum spin of node \( i \) in the graph. A downturn $\downarrow$ corresponds to $\sigma_{i} = -1$ and an upturn $\uparrow$ corresponds to $\sigma_{i} = +1$. The coefficients \( J_{ij} \) represent the strength of the interaction between nodes \( i \) and \( j \), while \( h_i \) denotes the external bias applied to node \( i \). The energy function \( H(\sigma) \) evaluates the total energy of a configuration (modeled with $\sigma = (\sigma_{1}, \sigma_{2} \ldots \sigma_{n})$) by adding the two-body interaction terms \( J_{ij} \sigma_i \sigma_j \) and the bias contributions \( h_i \sigma_i \).

% The Ising model has been directly applied in quantum computing, particularly in the context of quantum annealing. In D-Wave quantum computers, the nodes represent qubits with their spin $\sigma_{i}$ and the physical realization of this model is achieved through couplers that encode the interaction coefficients \( J_{ij} \) by entangling qubits \footnote{The definition of entanglement is further explain in section $5.1.7$} and through magnetic biases expressed through the \( h_i \) terms. Together, the biases and couplers define an energy landscape in which D-Wave quantum computers find the minimum energy during quantum annealing.

% Ising models and QUBO models can easily be mapped to each other through bijective mapping of QUBO binary variables to Ising spin variables. QUBO binary variables can be assigned to Ising spin variables using:
% \[\sigma_{i} = 2x_{i} -1\]
% and Ising spin variables can be mapped to QUBO binary variables using:
% \[x_{i} = \frac{\sigma_{i}+1}{2}\]

The simple physical implementation of Ising models and the direct correspondence between the Ising model and QUBO (Quadratic Unconstrained Binary Optimization) \footnote{Will further be discussed in \ref{subsec:qa}} motivates the use of QUBO formulations for framing optimization problems suitable for quantum computing platforms. 

% These quantum algorithms tackle problems using fundamentally different paradigms compared to classical methods, offering potential for both analytical insights and computational speedups. Although quantum heuristics are unlikely to provide exponential speed-ups to NP-Hard problems \citep{Preskill_2018}, they have demonstrated promise in delivering quadratic speed-ups and improved heuristic approaches for certain domains of problems \citep{Guerreschi_2019}.

\subsubsection{Measurements}
\label{subsubsec:measurements}
When a qubit is measured, its superposition is collapsed and deterministically measured in a $\ket{0}$ or $\ket{1}$ state. For example, when measuring $\ket{\psi}$,
\[\ket{\psi} = \alpha\ket{0} + \beta\ket{1},\]
it will come up as $\ket{0}$ with probability $|\alpha|^2$ and $\ket{1}$ with probability $|\beta|^2$. Similarly for $\ket{\phi}$,
\[\ket{\phi} = \frac{1}{\sqrt{2}}\ket{0} + \frac{1}{\sqrt{2}}\ket{1},\]
there is $\frac{1}{2}$ chance $\ket{\phi}$ is measured as $\ket{0}$, and $\frac{1}{2}$ chance $\ket{\phi}$ is measured as $\ket{1}$.

\noindent Although measurement gates are typically applied in the computational basis (i.e., using the basis states $\ket{0}$ and $\ket{1}$), they can be explicitly defined for any quantum state $\ket{\psi}$ through the following equation:
\[M_{\psi} = \ket{\psi}\bra{\psi}.\]
Note that $\bra{\psi}$ is the bra of $\psi$ and is the conjugate transpose of $\ket{\psi}$. Applying this to the computational basis states, $\ket{0}$ and $\ket{1},$ we get:
$$M_{\ket{0}}=\ket{0}\bra{0}=\begin{pmatrix}
        1 & 0 \\
        0 & 0
    \end{pmatrix},$$
$$M_{\ket{1}}=\ket{1}\bra{1}=\begin{pmatrix}
        0 & 0 \\
        0 & 1
    \end{pmatrix}.$$

%\noindent You can measure state $m$ in the basis state $\ket{b}$ with probability
%$$P(\ket{\psi})=\bra{\psi}M^{\dagger}_{\ket{b}}M_{\ket{b}}\ket{\psi}.$$

\subsubsection{Noisy Intermediate-Scale Quantum (NISQ)}
\label{subsubsec:NISQ}
In the realm of gate-based quantum computing, current technology is constrained to the Noisy Intermediate-Scale Quantum (NISQ) era. As indicated by the term ``NISQ," these processors are limited in both size and accuracy \citep{Brooks2019}. For example, IBM's Osprey, the most powerful publicly available quantum computer to date, has only 433 qubits, which restricts the range of problems it can solve. As of 2025, only two gate-based processors and three annealing-based processors have more than 1000 qubits.

Furthermore, the hardware's quantum capabilities are extremely sensitive. As mentioned earlier, qubits make up the core of quantum systems, and their superposition and entanglement properties offer quantum speedup. Unfortunately, current qubits are noisy and suffer from decoherence (loss of quantum properties) and quantum gate imperfections due to environmental influence. For these reasons, only circuits with low quantum depth (a few layers of quantum gates) have even moderate levels of accuracy. Although there is active research on promising error correction schemes, they are not yet applicable to larger-scale devices that prevent fault-tolerant computing. 

Once both of these hurdles are overcome, the post-NISQ era will see the emergence of fault-tolerant quantum computers with tens of thousands of qubits, capable of performing tasks far beyond the reach of classical systems. One such milestone is the ability to break RSA encryption using Shor’s algorithm—a task that would represent a definitive demonstration of quantum advantage. Leading experts estimate that this so-called "Q-Day"—the point at which quantum hardware can reliably run Shor’s algorithm to factor RSA-2048 keys—could arrive sometime after 2035~\citep{gidney2025rsa}.

\subsection{Foundations of Quantum Computing (Part II): Algorithms for QUBO Models}
\label{subsec:found2}

Adiabatic Quantum Computing (AQC) is an alternative form to gate-based that uses the adiabatic theorem.\footnote{The Adiabatic Theorem will be further explained in \ref{subsec:qa}} AQC uses the continuous evolution of a quantum system, and it is theoretically universal \citep{Albash_2018} as well. AQC is primarily implemented using quantum annealing. 
%, quantum annealers can only be used to solve sampling or optimization problems.

%Quantum annealers, like many gate-based computers, also use superconducting qubits. However, they operate these qubits in an analog fashion, in contrast to gate-based quantum computers, which manipulate qubits digitally. As a result, quantum annealers can scale to thousands of qubits, whereas gate-based quantum devices are currently limited to just over 1,000 qubits.

%Annealing algorithms are metaheuristic techniques for optimization problems inspired by the physical process of annealing in metallurgy, where metals are heated and then gradually cooled to remove defects and improve their properties.

To better appreciate the nuances of quantum annealing, we will first delve into simulated annealing. This classical technique will provide a foundation for understanding how its quantum counterpart builds on and improves upon these principles.

Quantum heuristics like the Quantum Approximate Optimization Algorithm (QAOA) and quantum annealing (QA) have demonstrated competitive performance against classical methods such as branch-and-cut for solving large-scale Quadratic Unconstrained Binary Optimization (QUBO) problems. For instance, a benchmarking study found that a hybrid quantum-classical solver achieved a relative accuracy of 0.013\% and solved problems with up to 10,000 variables approximately 6,561 times faster than the best classical solvers \citep{https://doi.org/10.48550/arxiv.2504.06201}. However, classical branch-and-cut methods, implemented in solvers like CPLEX and Gurobi, remain robust for smaller to medium-sized problems, providing exact solutions with well-understood performance metrics. In contrast, quantum annealing has shown advantages in certain problem instances, such as the Maximum-Cut problem, where it outperformed classical simulated annealing in terms of solution quality and computational time \citep{vodeb2024accuracyperformanceevaluationquantum}. Nevertheless, the performance of quantum heuristics can be sensitive to problem structure and hardware constraints, and they may not consistently outperform classical methods across all problem types. Therefore, while quantum heuristics offer promising avenues for optimization, classical methods like branch-and-cut continue to be valuable tools, especially for problems where exact solutions are critical.
\subsubsection{Simulated Annealing}
\label{subsec:sa}
Simulated annealing (SA) is a probabilistic algorithm designed to find heuristic solutions to optimization problems. The method draws inspiration from the physical annealing process by gradually reducing the ``temperature" of the system to locate a low-energy or optimal configuration.

SA begins by mapping the target problem to a cost function. This function maps possible solutions to their costs (also known as energies) to evaluate the quality of the solution. For the two solutions, the one with the lower energy is considered better.

The algorithm first initializes with a random solution and iteratively refines it to approximate a global optimum. In each iteration, the algorithm generates a neighboring solution by applying small perturbations to the current solution and evaluates its energy using the cost function. If the neighbor has a lower energy, it replaces the current solution. However, SA stands out from other metaheuristic methods by employing a \textbf{Temperature Schedule} and \textbf{Cooling Process} to determine whether to accept a neighboring solution, which makes it more resistant to getting stuck in local minima.
\paragraph{Temperature Schedule}
The likelihood of replacing the current solution with a neighbor of higher energy depends on the temperature $T$, which is initialized at the beginning of the algorithm. The probability of acceptance is given by:
\[
p(f, x, x', T) = \exp\left(-\frac{f(x') - f(x)}{T}\right),
\]
where $f(x')$ and $f(x)$ are the energies of the neighboring and current solutions, respectively. This acceptance probability is derived from thermal annealing and governs the exploration process. When $T$ is high or the energy difference $\Delta E := f(x') - f(x)$ is small, the algorithm is more likely to accept worse solutions, promoting exploration. Over time, $T$ decreases, reducing the probability of such moves and encouraging the exploitation of promising regions. A key limitation of SA is its reduced ability to explore solutions when $\Delta E$ is large.

\paragraph{Cooling Process}
The cooling process reduces the temperature $T$ at each iteration, typically using an exponential decay governed by a cooling factor $0 < \alpha < 1$:
\[
T_{\text{new}} = \alpha \cdot T_{\text{old}}.
\]
As $T$ decreases, the algorithm becomes less likely to accept suboptimal solutions, progressively shifting its focus to exploiting the current solution. This gradual reduction ensures the convergence to solutions with lower optimality error.

Simulated annealing offers several advantages, including its ability to escape local minima \citep{17235}, simplicity of implementation, and adaptability to various problem domains. However, it also has limitations, such as its dependence on parameter adjustment (e.g., initial temperature, cooling factor), slower convergence compared to other methods \citep{10.1137/S1052623497329683}, and suboptimal performance when there is a large energy difference ($\Delta E$) \citep{article}. A pseudocode for Simulated Annealing is provided below in Algorithm \ref{algo:sannealing}.
\begin{algorithm}
\caption{Simulated Annealing}
\begin{algorithmic}[1]
\Require Objective function $f(x)$, initial solution $x$, initial temperature $T_{\text{initial}}$, final temperature $T_{\text{final}}$, cooling factor $\alpha$, and maximum iterations per temperature $N_{\text{iter}}$.
\Ensure Approximate solution $x_{\text{best}}$ minimizing $f(x)$.

\Procedure{SimulatedAnnealing}{$f$, $x$, $T_{\text{initial}}$, $T_{\text{final}}$, $\alpha$, $N_{\text{iter}}$}
    \State $T \gets T_{\text{initial}}$ \Comment{Initialize temperature}
    \State $x_{\text{best}} \gets x$ \Comment{Initialize the best solution with the initial solution}

    \While{$T > T_{\text{final}}$} \Comment{Iterate until the system cools down}
        \For{$i = 1$ to $N_{\text{iter}}$} \Comment{Iterate for a fixed number of steps at each temperature}
            \State $x' \gets$ Generate a random neighbor of $x$ \Comment{Perturb the current solution}
            \State $\Delta E \gets f(x') - f(x)$ \Comment{Evaluate the energy difference}
            
            \If{$\Delta E < 0$} \Comment{If the neighbor is better, accept it}
                \State $x \gets x'$
                \If{$f(x) < f(x_{\text{best}})$} \Comment{Update the best solution if needed}
                    \State $x_{\text{best}} \gets x$
                \EndIf
            \Else
                \State $p_{\text{accept}} \gets e^{-\Delta E / T}$ \Comment{Calculate acceptance probability}
                \If{$\text{rand}(0, 1) < p_{\text{accept}}$} \Comment{Accept inferior solution probabilistically}
                    \State $x \gets x'$
                \EndIf
            \EndIf
        \EndFor
        \State $T \gets \alpha \times T$ \Comment{Reduce temperature based on the cooling factor}
    \EndWhile

    \State \textbf{return} $x_{\text{best}}$ \Comment{Return the best solution found}
\EndProcedure
\end{algorithmic}
\label{algo:sannealing}
\end{algorithm}

\subsubsection{Quantum Annealing}
\label{subsec:qa}
Quantum annealing (QA) represents a prominent example of adiabatic quantum computing and shares similarities with simulated annealing. It leverages quantum mechanical principles, particularly quantum tunneling, to potentially outperform classical algorithms. Unlike many other quantum computing techniques, quantum annealing is highly specialized for optimization problems, making it particularly effective for solving complex combinatorial optimization tasks. This method is implemented on quantum annealing platforms like D-Wave’s quantum processors, which are designed to find the minimum energy configuration in such optimization problems.

\paragraph{Mapping the Problem: QUBO and Ising Models}

To understand how quantum annealing works on these platforms, it's essential to explore the relationship between the Ising and QUBO models. The Ising model, introduced in \ref{subsubsec:ising}, plays a fundamental role in quantum computing, especially in quantum annealing. In the context of D-Wave's quantum computers, the nodes of the system represent qubits, which can be in a superposition of states with their associated spin variable \(\sigma_i\). The physical realization of the Ising model is achieved by coupling qubits, where interaction coefficients \(J_{ij}\) are encoded, and introducing magnetic biases, represented by the terms \(h_i\). These biases and couplings define the energy landscape that the system traverses during the quantum annealing process, where the goal is to find the ground state that corresponds to the optimal solution.

The relationship between the Ising and QUBO models is particularly important because quantum annealing can solve problems formulated in either model. The Ising model and the Quadratic Unconstrained Binary Optimization (QUBO) model are closely related and can be mapped to each other through a simple bijective transformation between their binary variables. Specifically,
\[
\sigma_i = 2x_i - 1 \text{ and }
x_i = \frac{\sigma_i + 1}{2}.
\]
This mapping enables quantum annealing to solve optimization problems expressed in either model, facilitating its application across various domains.

\paragraph{Adiabatic Theorem}

At the core of quantum annealing lies the \textbf{Adiabatic Theorem}, which states that a quantum system initially in the ground state of a time-dependent Hamiltonian will remain in its instantaneous ground state throughout the evolution, provided that the evolution is sufficiently slow and the Hamiltonian varies smoothly. This property--for a quantum mechanical system to stay in its current instantaneous state (also known as eigenstate)--is known as adiabaticity. In the context of quantum optimization, this process begins with the system in the ground state of a simple \emph{driver Hamiltonian} \( H_D \), and gradually transforms toward a \emph{cost Hamiltonian} \( H_C \) that encodes the optimization problem.

This transformation is governed by a time-dependent Hamiltonian of the form:
\[
H(s(t)) = (1 - s(t)) H_D + s(t) H_C,
\]
where \( s(t) \in [0,1] \) is a monotonically increasing scheduling function over time \( t \in [0, T] \), with \( s(0) = 0 \) and \( s(T) = 1 \).

A key quantity governing the success of this process is the \emph{spectral gap} \( \Delta(s) \), defined as the energy difference between the ground state and the first excited state of \( H(s) \). The \emph{minimum spectral gap} over the entire evolution,
\[
\Delta_{\min} = \min_{s \in [0,1]} \left[ E_1(s) - E_0(s) \right],
\]
determines how slowly the system must evolve to maintain adiabaticity.

To guarantee that the system remains in the ground state throughout the evolution, the total evolution time \( T \) must satisfy the adiabatic condition:
\[
T \gg \frac{1}{\Delta_{\min}^2},
\]
up to factors depending on the norms of the derivatives of \( H(s) \). If this condition is met, the system will end up in the ground state of \( H_C \) at \( t = T \), thus yielding the optimal solution to the encoded problem.

\paragraph{Quantum Annealing Workflow}

In quantum annealing on an $n$-qubit system, the optimization problem is first converted into an objective function, similar to the cost function used in simulated annealing. This function assigns energy values to each possible solution, with the lowest energy representing the optimal solution. This objective function is then encoded into a cost Hamiltonian that defines the energy landscape of the problem.

Quantum annealing utilizes two Hamiltonians: the \textbf{driver Hamiltonian} and the \textbf{cost Hamiltonian}.

\begin{itemize}
    \item \textbf{Driver Hamiltonian (\( H_D \)):} The driver Hamiltonian is designed to ensure that its ground state is easy to prepare, typically by putting all qubits in a uniform superposition. It is defined as:
    \[
    H_D = -\sum_{i=1}^{n} \sigma_x^{(i)},
    \]
    where \( \sigma_x^{(i)} \) denotes the Pauli X operator that acts on the \( i \)th qubit. This Hamiltonian is responsible for initializing the system and guiding it through the annealing process.
    
    \item \textbf{Cost Hamiltonian (\( H_C \)):} The cost Hamiltonian encodes the specific optimization problem. It represents the energy associated with different states of the quantum system, with the optimal solution corresponding to the ground state of \( H_C \). Finding the ground state of \( H_C \) is typically NP-Hard, making it difficult for classical algorithms to solve.
\end{itemize}

These two Hamiltonians (\( H_D \) and \( H_C \)) are combined in quantum annealing, using the adiabatic theorem to guide the system’s evolution from the ground state of \( H_D \) to the ground state of \( H_C \) over time.

The D-Wave annealer, for instance, receives and solves an Ising Hamiltonian that is a combination of the initial and final Hamiltonians. The Ising Hamiltonian is given by:

\[
H_{\text{Ising}} = -\frac{A(s)}{2} \underbrace{\left( \sum^{n}_{i=1} \sigma_X^{(i)} \right)}_{\text{Initial Hamiltonian}} + \frac{B(s)}{2} \underbrace{\left( \sum^{n}_{i=1} h_i \sigma_Z^{(i)} + \sum_{1 \leq j < i \leq n} J_{ij} \sigma_Z^{(i)} \sigma_Z^{(j)} \right)}_{\text{Final Hamiltonian}},
\]
where \( \sigma_i \) are the Pauli matrices acting on qubit \( q_i \), and \( h_i \) and \( J_{ij} \) represent the biases and couplings, respectively. As shown in the equation, the Ising Hamiltonian consists of an initial Hamiltonian (represented by the first sum of Pauli-X operators) and a final Hamiltonian (comprising biases and couplings involving Pauli-Z operators).

\paragraph{Quantum Annealing Process}
The Ising Hamiltonian begins with \( A(s) \) being a very large negative value and \( B(s) \) being a small positive value. As a result, the physical system starts in the ground state of the initial Hamiltonian (or in a uniform superposition). During the annealing process, \( A(s) \) gradually decreases in magnitude while \( B(s) \) increases. The magnetic biases and couplings guide the physical system towards the ground state of the problem Hamiltonian. At the end of the anneal, the qubits collapse to the ground-state energy of the problem Hamiltonian, which represents the optimal solution.

\begin{figure}
    \centering
    \includegraphics[width=0.9\linewidth]{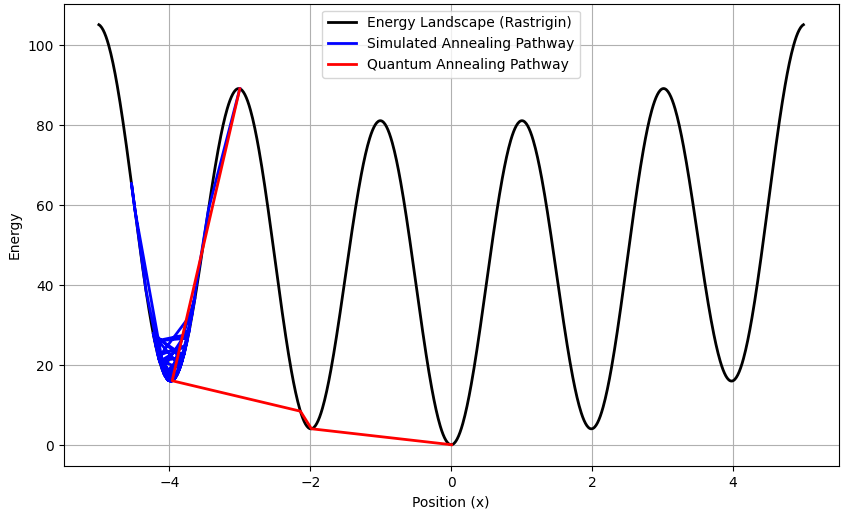}
    \caption{Simulated Annealing vs Quantum Annealing Optimization Pathway over a complex function.}
    \label{fig:savsqaperf}
\end{figure}

The adiabatic theorem is crucial to the success of quantum annealing, as it ensures that the final physical system is in its ground state (i.e., the optimal solution is obtained) if the evolution is slow enough.
\begin{algorithm}
\caption{Quantum Annealing}
\begin{algorithmic}[1]
\Procedure{quantum\_annealing}{$| \psi_0 \rangle$, $A(s)$, $B(s)$, $h_i$, $J_{ij}$, $T$}
  \State $s \gets 0$  \Comment{Initialize annealing parameter}
  \State $| \psi(s) \rangle \gets | \psi_0 \rangle$  \Comment{Initial state}
  \State $\text{best\_solution} \gets | \psi(s) \rangle$  \Comment{Track best solution}

  \While{$s < 1$}  \Comment{Loop until final annealing step}
    \State Compute $H_{\text{Ising}}(s)$:
    \[
    H_{\text{Ising}}(s) = -\frac{A(s)}{2} \sum_{i=1}^{n} \sigma_{X}^{(i)} + \frac{B(s)}{2} \left( \sum_{i=1}^{n} h_{i} \sigma_{Z}^{(i)} + \sum_{1 \leq j < i \leq n} J_{ij} \sigma_{Z}^{(i)} \sigma_{Z}^{(j)} \right)
    \]
    \State Evolve state $| \psi(s) \rangle$ under the Schrödinger equation for $\Delta s$ using $H_{\text{Ising}}(s)$:
    \[
    i \hbar \frac{d}{ds} | \psi(s) \rangle = H_{\text{Ising}}(s) | \psi(s) \rangle
    \]
    \State Update $s \gets s + \Delta s$
  \EndWhile
  
  \State $\text{best\_solution} \gets$ Measure the ground state $| \psi(s) \rangle$ \Comment{Final state corresponds to optimal solution}
  \State \textbf{return} $\text{best\_solution}$
\EndProcedure
\end{algorithmic}
\label{algo:qannealing}
\end{algorithm}
\paragraph{Quantum Tunneling}
During time evolution, the quantum system could encounter energy barriers where the differences in energy between the current state and neighboring states are very large. Simulated annealing often struggles with finding the global optimum in these circumstances, due to its acceptance probability function, but quantum annealing prevails due to Quantum Tunneling \citep{SHIFMAN_2002}. Because of the wavelike properties of the encoded Hamiltonians, there is a non-zero probability that the system can tunnel through the energy barrier avoiding local optima and find the global optimum (Figure \ref{fig:savsqaperf}). This phenomenon is the main quantum-mechanical advantage of QA over SA. A pseudocode for Quantum Annealing is provided in Algorithm \ref{algo:qannealing}.

\subsubsection{Quantum Approximate Optimization Algorithm (QAOA)}
\label{subsec:qaoa}
The Quantum Approximate Optimization Algorithm (QAOA) \citep{farhi2014quantum} is a quantum-classical hybrid algorithm designed for combinatorial optimization problems. Unlike other quantum algorithms, QAOA has a relatively low circuit depth, making it more resilient to quantum decoherence, which is a key challenge in near-term quantum computing. As a result, QAOA is considered one of the promising NISQ (Noisy Intermediate Scale Quantum) \ref{subsubsec:NISQ} algorithms and has potential applications in areas such as finance, operations research and logistics.

QAOA can be viewed as a trotterized (a form of discretized approximation) version of quantum annealing (QA). In quantum annealing, a system evolves adiabatically from a simple initial Hamiltonian (the driver Hamiltonian) to a problem Hamiltonian (the cost Hamiltonian). However, the analog process of quantum annealing relies on continuous-time evolution, which is not directly realizable on digital quantum computers. To adapt the process for gate-based quantum computers, \textbf{trotterization} is used to approximate continuous evolution in discrete steps, using alternating applications of cost and driver operators.

%\paragraph{Comparison to VQE}

%QAOA shares similarities with the Variational Quantum Eigensolver (VQE), which solves eigenvalue problems, particularly in computational chemistry. While VQE focuses on finding the ground state energy of molecules, QAOA is used for combinatorial optimization problems. Despite their differences in application, both algorithms share a similar variational structure.

\paragraph{Problem Mapping}

The first step in implementing QAOA is to map the combinatorial optimization problem to an objective function \( f \), which quantifies how far a given candidate solution is from the optimal one. This objective function is analogous to the cost function used in simulated annealing and quantum annealing. In QAOA, the goal is to minimize the error function \( f \) by iterating through potential solutions in a probabilistic way.

Since quantum circuits operate probabilistically, QAOA evaluates the error function \( f \) by computing its expected value over a set of potential solutions. The expected value of the error function corresponds to the quality of the solution, where lower values indicate better solutions.

The algorithm initializes a set of parameters \( \{\beta_i\}_{i \geq 0} \) and \( \{\gamma_i\}_{i \geq 0} \), which control the evolution of the quantum circuit. These parameters are then fed into the quantum circuit, which produces an initial quantum state that is evaluated using the objective function \( f \).

\paragraph{Creating the Parametrized Circuit through Trotterization}

QAOA constructs a parametrized quantum circuit composed of a Hadamard gate layer, followed by alternating layers of mixers and cost operators, which work together to explore the solution space and guide the system toward the optimal solution.

%Unlike quantum annealing, which continuously evolves the quantum state, QAOA uses trotterization to discretize the evolution process into small, discrete time steps. Trotterization approximates the continuous evolution of the system by applying discrete unitary operations. These unitary operations define the cost and mixer operators in the quantum circuit.

\textbf{1. Initial State Preparation with Hadamard Gates}

The quantum circuit begins by applying Hadamard gates to the initial qubits to transform them from the \( |0\rangle \) state into a superposition of all possible states on the $n$ qubits:
\[
|+\rangle := \frac{1}{\sqrt{2}} |0\rangle + \frac{1}{\sqrt{2}} |1\rangle,
\]
\[
\ket{\psi(0)} = \ket{+}^{n}.
\]
This step ensures that the quantum system explores all possible configurations of the optimization problem.

\textbf{2. Cost and Driver Operators in QAOA through Trotterization}

Similarly, assume that we have an $n$-qubit system. In QAOA, the evolution of the quantum state is governed by two key operators: the cost operator and the mixer operator. These operators are derived from the previously mentioned cost Hamiltonian and driver Hamiltonian described in the quantum annealing section.

In traditional quantum annealing, the problem is encoded in a cost Hamiltonian \( H_C \) and a driver Hamiltonian \( H_D \), which together define the total Hamiltonian \( H = H_C + H_D \). The quantum state \( \ket{\psi} \) evolves according to the Schrödinger equation:
\[
i \hbar \frac{d}{dt} \ket{\psi(t)} = H \ket{\psi(t)}.
\]
This evolution leads to the solution:
\[
\ket{\psi(t)} = e^{-i H t} \ket{\psi(0)}.
\]
In QAOA, we discretize the continuous-time evolution described in quantum annealing \ref{subsec:qa} using Trotterization. This allows us to approximate the evolution using small, discrete steps, represented by the cost operator and the mixer operator \footnote{The construction of these operators for arbitrary QUBOs is described in \ref{subsec:qai}}. For this reason, the Hamiltonians described in this section are also the same as those described in \ref{subsec:qa}.

\paragraph{Cost Hamiltonian (Optimization Problem Encoding)}

The cost Hamiltonian \( H_C \)  encodes the objective function and is given by:
\[
H_C = \sum_{i=1}^{n} h_i \sigma_z^{(i)} + \sum_{1\leq j < i \leq n} J_{ij} \sigma_z^{(i)} \sigma_z^{(j)},
\]
where \( \sigma_z^{(i)} \) is the Pauli-Z operator acting on the qubit \( i \), and \( h_i \) and \( J_{ij} \) are the biases and coupling constants. The goal is to find the ground state of \( H_C \), which corresponds to the optimal solution.

Trotterization approximates the cost operator as:
\[
e^{-i \gamma H_C} \approx \prod_{1 \leq j < i \leq n} e^{-i \gamma J_{ij} \sigma_z^{(i)} \sigma_z^{(j)}} \prod_{i=1}^{n} e^{-i \gamma h_i \sigma_z^{(i)}}.
\]
\paragraph{Driver Hamiltonian (Mixer Operator)}

The driver Hamiltonian \( H_D \) encourages the system to explore different possible states. It typically takes the form of the Pauli-X operator:
\[
H_D = - \sum_i \sigma_x^{(i)},
\]
which mixes the basis states to prevent the system from getting stuck in local minima. The evolution under the driver Hamiltonian is discretized as:
\[
e^{-i \beta H_D} \approx \prod_{i} e^{-i \beta \sigma_x^{(i)}}.
\]
\paragraph{Combining the Cost and Driver Operators in QAOA}

In QAOA, the quantum state evolves by alternating between applying the cost operator and the driver operator. The total evolution after \( p \) layers is given by:
\[
|\psi(\beta, \gamma)\rangle = e^{-i \gamma_1 H_C} e^{-i \beta_1 H_D} e^{-i \gamma_2 H_C} e^{-i \beta_2 H_D} \dots e^{-i \gamma_p H_C} e^{-i \beta_p H_D} |\psi(0)\rangle.
\]
The parameters \( \beta_i \) and \( \gamma_i \) are classically optimized to minimize the expected value of the cost Hamiltonian, guiding the quantum system toward the optimal solution.

These operations are repeated for \( p \) layers, with \( p \) controlling the number of alternating steps between the cost and driver operators. The fully parametrized circuit can be expressed as
\[
|\psi\rangle = \prod_{i=1}^{p} e^{-i\beta_i H_D} e^{-i\gamma_i H_C} |\psi(0)\rangle,
\]
where \( |\psi\rangle \) is the quantum state after the \( p \) layers of operation. The circuit is called an ansatz and is parametrized by the \( 2p \) parameters \( \gamma \) and \( \beta \).

\paragraph{Classical Optimization}

Once the quantum circuit is prepared, a classical optimizer adjusts the parameters \( \{\beta_i\}_{i \geq 0} \) and \( \{\gamma_i\}_{i \geq 0} \) to minimize the expected value of the error function \( f \). Since the expectation is generally non-convex, classical optimization techniques, such as heuristics or gradient-based methods, are used to iteratively improve the parameter set.

After optimization is complete, the quantum circuit is measured and the result provides an approximate solution to the optimization problem.

A full pseudocode for QAOA is provided in Algorithm \ref{algo:qaoa}.

\begin{algorithm}[h]
\setlength{\baselineskip}{15pt}  % Adjust the vertical space between lines
\caption{Quantum Approximate Optimization Algorithm (QAOA)}
\begin{algorithmic}[1]
\Procedure{QAOA}{cost Hamiltonian $H_C$, Mixing Hamiltonian $H_M$, Number of Cycles $p$, Initial Parameters $\vec{\beta}, \vec{\gamma}$}
    \State Initialize state $| \psi \rangle = |+ \rangle^n$  \Comment{Start with all qubits in an equal superposition state}
    \State Initialize parameters $\vec{\beta}, \vec{\gamma} \in \mathbb{R}^p$ randomly  \Comment{Set initial values for parameters}
    
    \While{not converged}
        \For{$i = 1$ to $p$}  \Comment{Loop over each cycle}
            \State Apply the mixer Hamiltonian: $e^{-i \beta_i H_M}$  \Comment{Use $R_X$ gates to implement}
            \State Apply the cost Hamiltonian: $e^{-i \gamma_i H_C}$  \Comment{Use controlled ops. or $U_{ZZ}$ gates to implement}
        \EndFor
        
        \State Measure the state $| \psi \rangle$ to estimate the cost function value  \Comment{Evaluate the solution's quality}
        
        \State Update the parameters $\vec{\beta}, \vec{\gamma}$ using Classical Optimizer  \Comment{e.g., gradient descent to minimize cost}
    \EndWhile
    
    \State Measure the final state $| \psi \rangle$  \Comment{Obtain the final result after convergence}
    \State \textbf{return} Measurement outcome (approximate solution)  \Comment{Return the best solution found}
\EndProcedure
\end{algorithmic}
\label{algo:qaoa}
\end{algorithm}

The means to obtain mixer and cost Hamiltonians for specific QUBOs by constructing mixer and cost operators will be discussed in more detail in Section \ref{sec:solve}.

\subsubsection{Summary}
\label{subsec:summary}
\begin{figure}[h]
\centering
\begin{subfigure}[t]{0.45\textwidth}
    \centering
    \includegraphics[width=\linewidth]{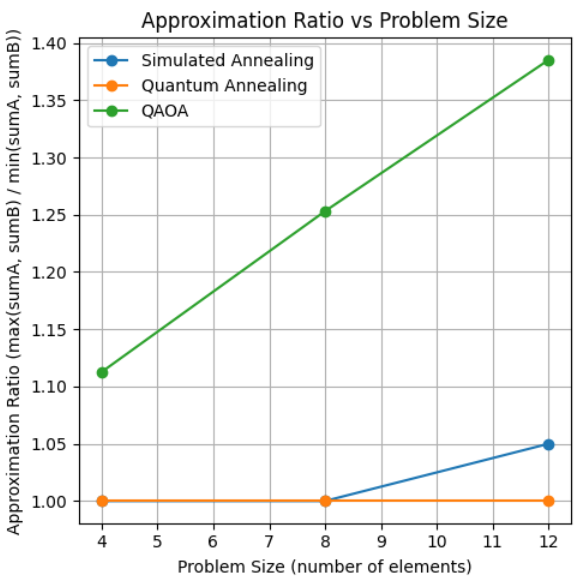}
    \caption{Quality of solutions versus input size.}
    \label{fig:approx_prob_size}
\end{subfigure}%
\hspace{0.05\textwidth}
\begin{subfigure}[t]{0.45\textwidth}
    \centering
    \includegraphics[width=\linewidth]{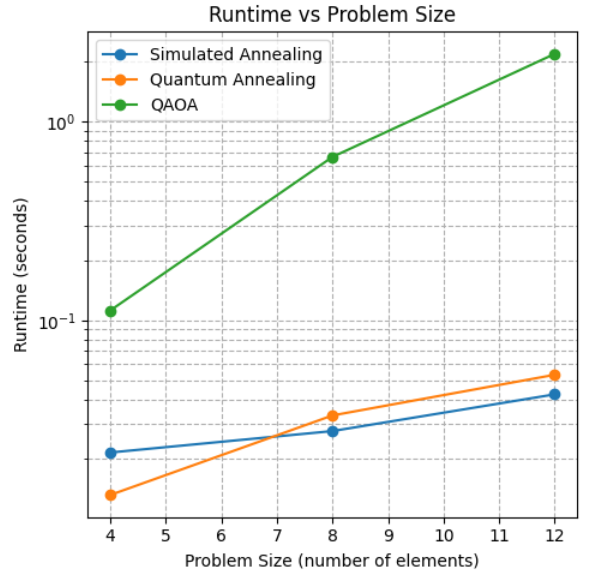}
    \caption{Computational runtime versus input size.}
    \label{fig:runtimesize}
\end{subfigure}
\caption{Performance comparison of SA, QA, and QAOA on the number partitioning problem.}
\label{fig:performance_analysis}
\end{figure}

\noindent
We summarize and compare the three algorithms (SA, QA and QAOA) in Table~\ref{fig:annealing_comparison} along several attributes. Figure~\ref{fig:performance_analysis} shows Approximation Ratios (left) and Runtimes (right) for small instances of the Number Partitioning problem. SA and QA are better than QAOA on both metrics. SA and  QA are competitive for  small problems. As problem size increases further (not shown here), QAOA falls further behind; QA reaches its limit due to hardware size, connectivity and fidelity.

\begin{figure}[h]
\centering
% TikZ Diagram
\begin{tikzpicture}[
  node distance=0.8cm and 3.2cm,
  every node/.style={font=\small, align=center},
  box/.style={rectangle, draw=black, rounded corners, fill=gray!10, minimum height=1cm, text width=3cm, align=center, thick},
  arrow/.style={-Stealth, thick, red}
]
\node[box] (sa) {Simulated Annealing (SA)};
\node[box, right=of sa] (qa) {Quantum Annealing (QA)};
\node[box, right=of qa] (qaoa) {QAOA};
\draw[arrow] (sa) -- node[above]{Quantum analog} (qa);
\draw[arrow] (qa) -- node[above]{Trotterized version} (qaoa);
\end{tikzpicture}

\vspace{1em}

% Table
\renewcommand{\arraystretch}{1.3}
\captionsetup{type=figure}
\begin{tabularx}{\textwidth}{|>{\raggedright\arraybackslash}p{3.3cm}|
                                >{\raggedright\arraybackslash}X|
                                >{\raggedright\arraybackslash}X|
                                >{\raggedright\arraybackslash}p{3.3cm}|}
\hline
\textbf{Attribute} & \textbf{Simulated Annealing (SA)} & \textbf{Quantum Annealing (QA)} & \textbf{QAOA} \\
\hline
\textbf{Type} & Classical metaheuristic optimization algorithm & Analog quantum computation using adiabatic evolution & Gate-based hybrid quantum-classical algorithm \\
\hline
\textbf{Hardware} & Classical processors (CPU/GPU) & Quantum annealers (e.g., D-Wave) & NISQ gate-model devices (e.g., IBM Q, IonQ) \\
\hline
\textbf{Search Mechanism} & Thermal fluctuations enable transitions over energy barriers & Quantum tunneling enables transitions through energy barriers & Quantum interference via parametrized unitary evolution \\
\hline
\textbf{Control Parameter} & Temperature schedule controlling randomness & Annealing schedule controlling interpolation between Hamiltonians & Optimized variational parameters (angles) for gates \\
\hline
\textbf{Scalability} & Scales to large problems classically, but convergence may be slow & Limited by coherence time, noise, and qubit connectivity & Limited by circuit depth, noise, and number of measurements \\
\hline
\textbf{Speed} & Slow convergence near local minima & Depends on minimum spectral gap (can be exponential) & Depends on depth $p$ and optimizer performance \\
\hline
\end{tabularx}
\caption{Comparison of Simulated Annealing (SA), Quantum Annealing (QA), and QAOA in terms of conceptual structure and algorithmic features.}
\label{fig:annealing_comparison}
\end{figure}

\section{Solving QUBOs on Quantum Computers}
\label{sec:solve}
Three algorithms for five previously mentioned problems are implemented in the following \href{https://github.com/arulrhikm/Solving-QUBOs-on-Quantum-Computers.git}{repository}: Quantum Approximate Optimization Algorithm (QAOA) \citep{farhi2014quantum}, Quantum Annealing (QA) \citep{article}, and simulated annealing (SA) \citep{doi:10.1126/science.220.4598.671}. QAOA is implemented in three different ways in IBM quantum back-ends and simulators: Vanilla QAOA (\href{https://github.com/arulrhikm/Solving-QUBOs-on-Quantum-Computers/blob/main/Notebook%201%20-%20Custom%20QAOA.ipynb}{Notebook 1}), OpenQAOA (\href{https://github.com/arulrhikm/Solving-QUBOs-on-Quantum-Computers/blob/main/Notebook%202%20-%20OpenQAOA%20QAOA.ipynb}{Notebook 2}), and Qiskit (\href{https://github.com/arulrhikm/Solving-QUBOs-on-Quantum-Computers/blob/main/Notebook%203%20-%20Qiskit%20Optimization%20QAOA.ipynb}{Notebook 3}). In this paper, we provide code snippets for implementing the Quantum Approximate Optimization Algorithm (QAOA) using IBM's QASM Simulator. For readers interested in applying QAOA to actual IBM quantum processors, we also provide similar implementations for the IBM backends ``IBM Kyiv," ``IBM Brisbane," and ``IBM Sherbrooke," all of which use the Eagle r3 Processor with 127 qubits. Specifically, these IBM machines and simulators were chosen due to their ease of access. Larger IBM machines or machines from other organizations like Rigetti or Google required paid access, which we avoid for the sake of this tutorial. The coding techniques work similarly across platforms due to OpenQAOA and Qiskit compatibility. Simulated annealing (\href{https://github.com/arulrhikm/Solving-QUBOs-on-Quantum-Computers/blob/main/Notebook%204%20-%20DWAVE%20Simulated%20Annealing.ipynb}{Notebook 4}) and quantum annealing (\href{https://github.com/arulrhikm/Solving-QUBOs-on-Quantum-Computers/blob/main/Notebook%205%20-%20DWAVE%20Quantum%20Annealing.ipynb}{Notebook 5}) are implemented (on D-Wave) using the API from Ocean SDK. For brevity, in each of the five subsections, we detail only one of the five problems for each algorithm, thereby covering all the problems and algorithms without repetition\footnote{The notebooks can be accessed in the following github: https://github.com/arulrhikm/Solving-QUBOs-on-Quantum-Computers.git}. We provide a brief description of each implementation's pros and cons at the end of this section.

\subsection{Vanilla QAOA}
\label{subsec:vqaoa}
 QAOA requires a cost Hamiltonian and a mixer Hamiltonian. These Hamiltonians are combined to create a QAOA circuit. All features were implemented following the instructions provided by the Qiskit textbook \citep{Qiskit-Textbook}. This first example focuses on the Number Partitioning problem to illustrate the steps and the code.
 
%The first algorithm that will be discussed is the Quantum Approximate Optimization Algorithm (QAOA). This first section will go over its vanilla implementation for the Number Partitioning Problem. \newline

%Although this code is the hardest to set up and least refined, it provides the highest customizability for solving QUBOs to QAOA. Unlike Qiskit and OpenQAOA SDKs which require the user to choose from a list of cost Hamiltonians, mixer Hamiltonians, optimizers, etc., this vanilla implementation allows the user to custom-define all these aspects. While this implementation uses Qiskit features for the basic circuit creation process, it does not utilize any of the built-in circuits that Qiskit Optimization provides. Thus this implementation is best when trying to modify mixer Hamiltonians or experiment with optimizers like Ant Colony Swarm Optimization. 

\subsubsection{Constructing Operators}
To create the cost and mixer operators, we need the tunable parameters $\gamma, \beta$ and the cost and mixer Hamiltonians of the problem. In the following section, we will explain how to obtain each of these structures and how to use them to create the corresponding operator.

First, note that the parameters $\gamma, \beta$ are arbitrarily initialized. In this example, they are each initialized as $0.5$. These parameters will be adjusted during the optimization phase for the QAOA circuit. 

Second, we aim to obtain the Hamiltonians. For an arbitrary QUBO in the form of
\[\sum_{i, j = 1}^{n}x_{i}Q_{ij}x_{j}+\sum_{i=1}^{n}c_{i}x_{i},\]
we get the corresponding cost Hamiltonian as
\[
H_{C} = \sum_{i,j=1}^{n} \frac{1}{4} Q_{ij} Z_i Z_j - \sum_{i=1}^{n} \frac{1}{2} \left( c_i + \sum_{j=1}^{n} Q_{ij} \right) Z_i,
\]
so the corresponding cost operator is
\[e^{-i\gamma H_{C}} = \prod_{i, j=1}^{n}R_{Z_{i}Z_{j}}\left(\frac{1}{4}Q_{ij}\gamma\right)\prod_{i=1}^{n}R_{Z_{i}}\left(\frac{1}{2}\left(c_{i} + \sum_{j=1}^{n}Q_{ij}\right)\gamma\right).\]
On a circuit with \( n \) qubits (circuit width of \( n \)-qubits), this operator can be constructed by applying two layers of gates. The first layer is a layer of \( R_Z \) gates to each qubit \( i \) (for \( i = 1, 2, \ldots, n \)), with angles \( \theta_{ij1} \):
\[\theta_{ij1} = \frac{\gamma}{2}(c_{i}+\sum_{i=1}^{n}Q_{ij}).\]
This layer of $R_{Z}$ gates is followed by a layer of $U_{ZZ}$ gates applied to each pair of qubits $i$ and $i$ at angles $\theta_{i2}$:
\[\theta_{i2} = \frac{1}{4}Q_{ij}\gamma.\]

\begin{lstlisting}[style=mystyle]
def cost_H(gammas, quadratics, linears):
  qc = QuantumCircuit(num_qubits, num_qubits)
  for i in range(len(linears)):
    qc.rz(1/2*(linears[(0, i)]+sum(quadratics[(i, j)] for j in range(num_qubits)))*gammas, i)

  for (i, j) in quadratics.keys():
    if i!=j:
      qc.rzz((1/4)*quadratics[(i, j)]*gammas, i, j)

  qc.barrier()
  return qc
\end{lstlisting}
\vspace{-0.9cm}
\begin{figure}[h]
    \centering
    \includegraphics[width=0.9\linewidth]{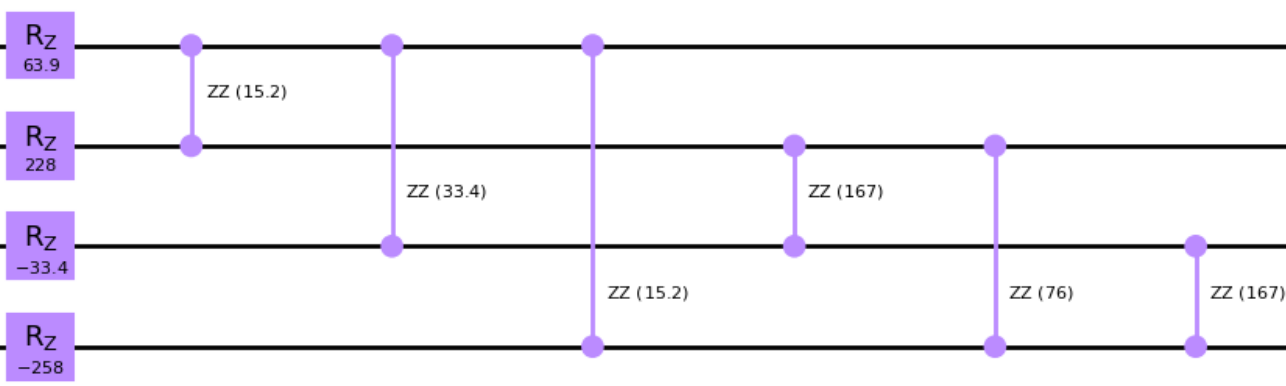}
\end{figure}
\vspace{-0.8cm}
Similarly, the mixer operator depends on $\beta$ and the mixer Hamiltonian.

The mixer Hamiltonian is
\[H_{M} = \sum_{i=1}^{n}\sigma_{X_{i}},\]
so the corresponding mixer operator is
\[e^{-i\beta H_{M}} = \prod_{i=1}^{n}R_{X}(2\beta).\]

This mixer operator can be applied by applying $R_{X}$ gates to each qubit \( i \) (for \( i = 1, 2, \ldots, n \)) at angle $\theta_{i3}$:
\[\theta_{i3} = 2\beta.\]
% The code and the resulting circuit (with cost and mixer Hamiltonians) are shown below:
\begin{lstlisting}[style=mystyle]
def mixer_H(betas):
    qc = QuantumCircuit(num_qubits, num_qubits)
    for i in range(num_qubits):
        qc.rx(2*betas, i)
    qc.barrier()
  return qc
  \end{lstlisting}
\vspace{-0.8cm}
\begin{figure}[h]
     \centering
     \includegraphics[width=0.9\linewidth]{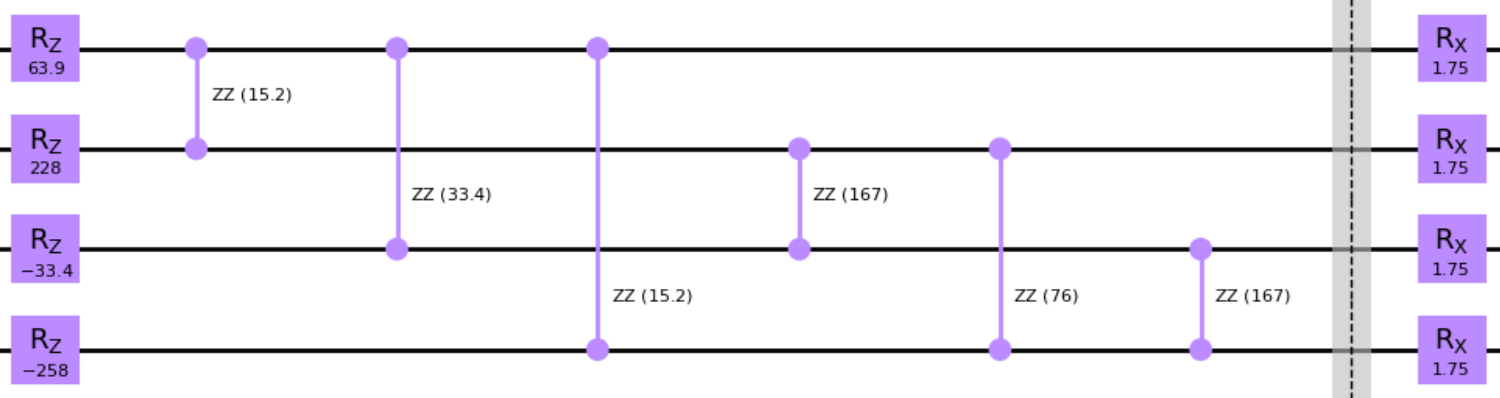}
\end{figure}
\subsubsection{QAOA Circuit}
% \begin{lstlisting}[style=mystyle]
%   def circuit(gammas, betas, quadratics, linears):
%     circuit = QuantumCircuit(num_qubits, num_qubits)
%     assert(len(betas) == len(gammas))
%     p = len(betas)

%     for i in range(num_qubits):
%         circuit.h(i)
%     circuit.barrier()

%     for i in range(p):
%         circuit &= cost_H(gammas[i], quadratics, linears)
%         circuit &= mixer_H(betas[i])

%     circuit.measure(range(num_qubits), range(num_qubits))

%     return circuit
% \end{lstlisting}
% \begin{figure}[h]
%     \centering
%     \includegraphics[width=1\linewidth]{CodePics/qiskit-qaoa-circuit.png}
% \end{figure}
\noindent Once the generic framework is established, the quadratic and linear coefficients of the Number Partitioning QUBO are substituted. The model is created using DOcplex \citep{cplex2009v12}, converted to a QUBO, and partitioned into quadratic and linear terms for gate implementation. The test example uses the set \(\{1, 5, 11, 5\}\) for partitioning.

\begin{lstlisting}[style=mystyle]
def circuit(gammas, betas, quadratics, linears):
  circuit = QuantumCircuit(num_qubits, num_qubits)
  p = len(betas)

  for i in range(num_qubits):
    circuit.h(i)
  circuit.barrier()

  for i in range(p):
    circuit &= cost_H(gammas[i], quadratics, linears)
    circuit &= mixer_H(betas[i])

  circuit.measure(range(num_qubits), range(num_qubits))

  return circuit    
\end{lstlisting}
\vspace{-0.5cm}
\noindent The following code generates the number partitioning instance, converts it to the QUBO model, and then collects the parameters of the QUBO model to construct the final circuit (with $p = 1$ for conciseness).
\begin{lstlisting}[style=mystyle]
arr = [1, 5, 11, 5]
n = len(arr)
c = sum(arr)

model = Model()
x = model.binary_var_list(n)
H = (c - 2*sum(arr[i]*x[i] for i in range(n)))**2
model.minimize(H)
problem = from_docplex_mp(model)

converter = QuadraticProgramToQubo()
qubo = converter.convert(problem)

quadratics_coeffs = qubo.objective.quadratic.coefficients
linears_coeffs = qubo.objective.linear.coefficients
constant = qubo.objective.constant
num_qubits = qubo.get_num_vars()
\end{lstlisting}
\vspace{-0.8cm}
\begin{figure}[h]
    \centering
    \includegraphics[width=\linewidth]{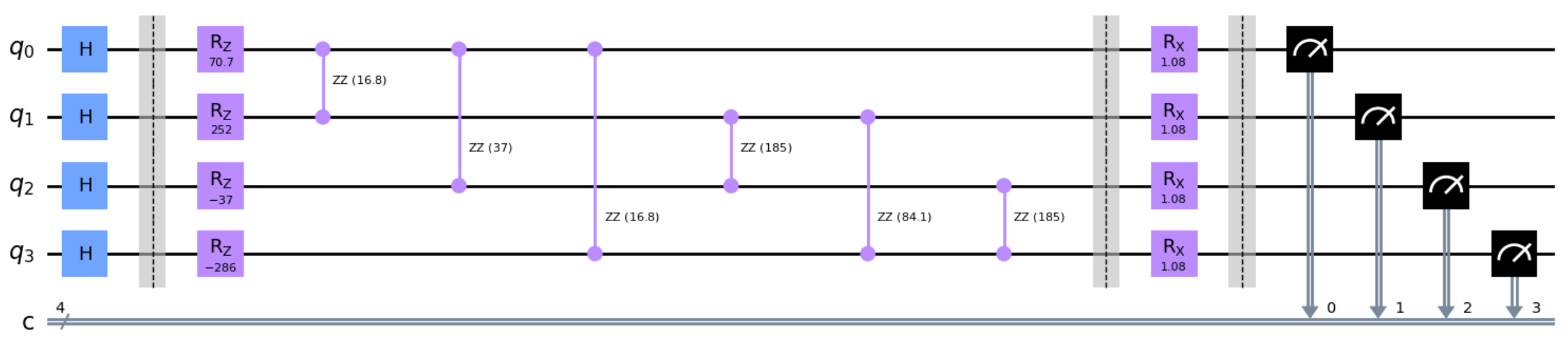}
\end{figure}
\noindent Once the circuit is created, its parameters $\gamma$ and $\beta$ are optimized using a classical optimizer. To optimize, we define an objective function and an expectation function for the output of the circuit. The objective function measures the quality of the QAOA output solutions, while the expectation function averages these solutions. The optimizer adjusts the parameters to minimize the expected value.
\begin{lstlisting}[style=mystyle]
def npp_obj(str):
  sum_0 = 0
  sum_1 = 0
  for i in range(len(str)):
    if str[i] == '0':
      sum_0 += arr[i]
    else:
      sum_1 += arr[i]
  return abs(sum_0-sum_1)
\end{lstlisting}
\begin{lstlisting}[style=mystyle]
def npp_expectation(thetas):
    backend = Aer.get_backend('qasm_simulator')
    gammas = theta[:int(len(thetas)/2)]
    betas = theta[int(len(thetas)/2):]
    pqc = circuit(gammas, betas, quadratics, linears)
    counts = execute(pqc, backend, shots=1000).result().get_counts()
    best_sol = max(counts, key=counts.get)

    return npp_obj(best_sol)    
\end{lstlisting}
% \begin{figure}[h]
%   \begin{subfigure}{0.4\textwidth}
%     \centering
% \includegraphics[width=0.6\textwidth]{CodePics/objectivefunction.png}
%   \caption{Number Partitioning objective function}
    
%   \end{subfigure}
%   \begin{subfigure}{0.6\textwidth}
%     \centering
% \includegraphics[width=\textwidth]{CodePics/expectationfunction.png}
%   \caption{Number Partitioning Expectation Function}
%   \label{fig:example7}
%   \end{subfigure}
%   \caption{QAOA Circuit Construction}
% \end{figure}
\vspace{-0.4cm}
\noindent After the optimization process is completed, the optimal parameters are passed to the final QAOA circuit and sampled from $1000$ shots. The number of shots can be adjusted; however, more shots will lead to more accurate results. 
\begin{figure}[h]
    \centering
    \includegraphics[width=0.6\linewidth]{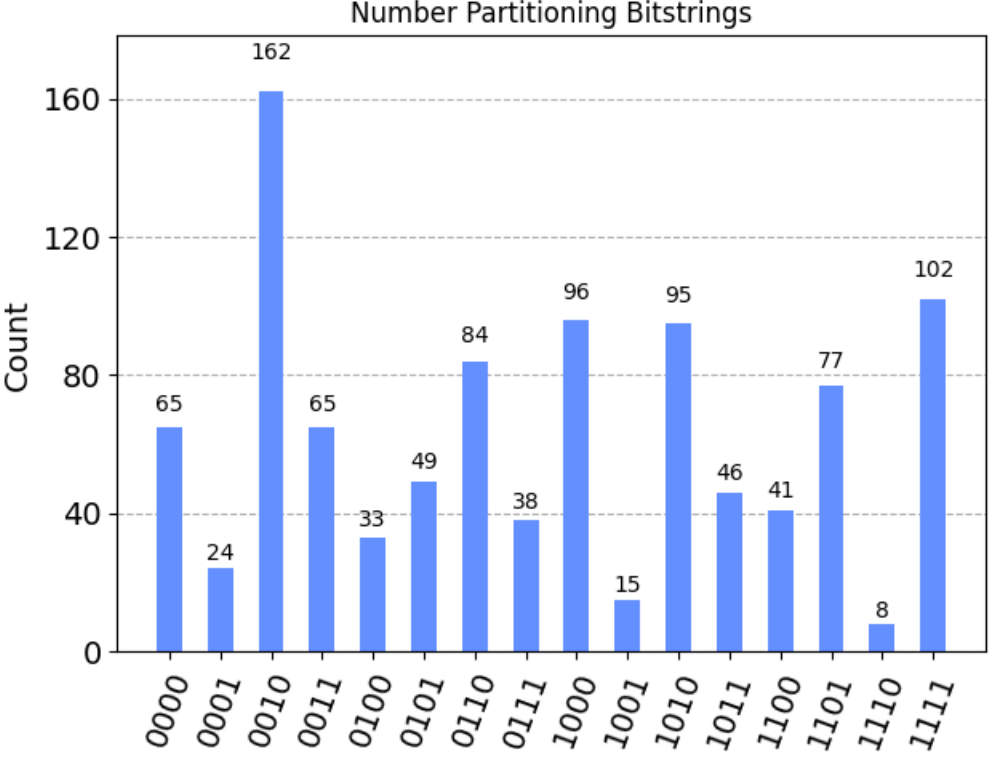}
    \caption{Sampled QAOA results for the Number Partitioning QUBO.}
    \label{fig:npp-hist}
\end{figure}
\noindent The resulting bitstrings correspond to the partition of the original array. For a string output $\texttt{s}$ and an original array $\texttt{arr}$, if $\texttt{s[i]}$ is $1$, $\texttt{arr[i]}$ is placed in the set $A$; if $\texttt{s[i]}=0$, it is in $S \setminus A$. In the histogram of Figure \ref{fig:npp-hist}, the most probable outcome is (0,0,1,0):  $A = \{11\}$ and $\{1, 5, 5\} = S \setminus A$.

\subsection{OpenQAOA}
\label{subsec:oqaoa}
 OpenQAOA \citep{sharma2022openqaoa} is a multi-backend SDK used to easily implement QAOA circuits. It provides simple yet very detailed implementations of QAOA circuits that can run on IBMQ devices and simulators, Rigetti Cloud Services \citep{Karalekas_2020}, Amazon Braket \citep{braket}, and Microsoft Azure \citep{Hooyberghs2022}. The parent company of OpenQAOA also provides custom simulators. 
 %Note that for resources like Rigetti, Braket, and Azure, the devices are only available with a subscription to their quantum services. 
%OpenQAOA is one of the most convenient SDKs available not only because of the large range of applicable devices but also the level of customizability and relevancy to ongoing research. OpenQAOA not only offers standard QAOA circuits but also recursive implementations RQAOA, as well as varied parametrization, initialization, and mixing strategies. OpenQAOA also offers a larger selection of optimizers than Qiskit-Optimization which are classified into three main categories: gradient-based, gradient-free, and shot-adaptive. Finally, it is easy to plot data like optimization pathways or bitstring distributions using OpenQAOA. All these capabilities and workflows can be further explored in Notebook 2.

\begin{figure}[h]    
     \centering
     \includegraphics[width=0.4\linewidth]{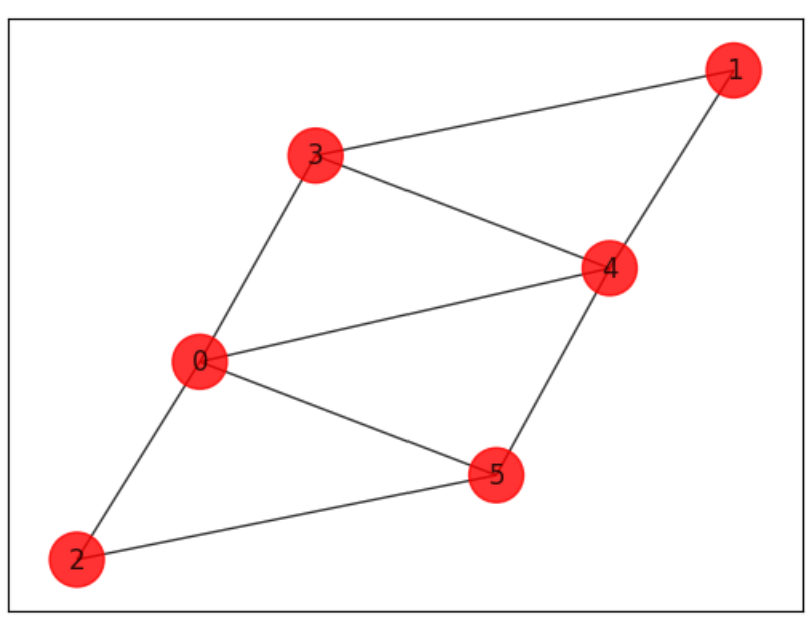}
     \caption{Generated Max-Cut Graph Instance.}
     \label{fig:maxcut-inst}
\end{figure}

%Although OpenQAOA offers more optimizers than Qiskit-Optimization, their list is still limited, and it is difficult to use custom optimizers. Furthermore similar to Qiskit-Optimization although you can adjust the circuit properties, you cannot directly modify the circuits also hindering the level of control.
We illustrate the implementation of the Max-Cut problem. This process requires the same steps as for the vanilla implementation of QAOA. However, most of the technical challenges are solved by the OpenQAOA API. Like the vanilla implementation, the first step is to create a problem instance (see Figure \ref{fig:maxcut-inst} for a generated graph instance) and then implement the QUBO model for it. The graph instance creation and the QUBO creation implementations are shown below:
% \begin{figure}[h]
%     \centering
%     \includegraphics[width=\linewidth]{CodePics/create_Max_Cut_Code.png}
%     \caption{Code for Max-Cut Problem Instance}
    
%     \centering
%     \includegraphics[width=0.5\linewidth]{CodePics/maxcutgraph.png}
%     \caption{Max-Cut Graph}
%     \label{fig:graph-mc}
% \end{figure}
\begin{lstlisting}[style=mystyle]
def draw_graph(G, colors, pos):
    default_axes = plt.axes()
    nx.draw_networkx(G, node_color=colors, node_size=600, alpha=0.8, ax=default_axes, pos=pos)
    edge_labels = nx.get_edge_attributes(G, "weight")
\end{lstlisting}
\begin{lstlisting}[style=mystyle]
G = nx.generators.fast_gnp_random_graph(n=6, p=0.5)
n = 6
colors = ["r" for node in G.nodes()]
pos = nx.spring_layout(G)
draw_graph(G, colors, pos)
\end{lstlisting}

The QUBO creation process for Max-Cut is very similar to the process with Number Partitioning.

\begin{lstlisting}[style=mystyle]
model = Model()
x = model.binary_var_list(n)
H = sum(2*x[e[0]]*x[e[1]] - x[e[0]] - x[e[1]] for e in G.edges)
model.minimize(H)

# Converting the Docplex model into its qubo representation
qubo = FromDocplex2IsingModel(model)

# Ising encoding of the QUBO problem
maxcut_ising = qubo.ising_model
\end{lstlisting}

\subsubsection{OpenQAOA Circuit}
Once the QUBO is created, OpenQAOA provides a short and easy implementation to create the QAOA circuit. To start, you only need to initialize the QAOA object (\texttt{q = QAOA()}) and then define its properties. These properties include whether the circuit will run locally on a simulator or virtually on a simulator or real device, the configurations of the cost and mixer operators, the number of repetitions of the operators, the classical optimizer and its properties, the number of shots to simulate the circuit, and much more. Once all the characteristics of the circuit are determined, the user can compile and optimize the parameters of the circuit.

\begin{lstlisting}[style=mystyle]
# initialize model
q = QAOA()

# device
q.set_device(create_device('local', 'vectorized'))

# circuit properties
q.set_circuit_properties(p=2, param_type='standard', init_type='rand', mixer_Hamiltonian='x')

# backend properties
q.set_backend_properties(n_shots = 1000)

# set optimizer and properties
q.set_classical_optimizer(method='vgd', jac="finite_difference")

q.compile(maxcut_ising)

q.optimize()
\end{lstlisting}
\subsubsection{Analysis Tools}
\begin{figure}[h]
  \centering
  \begin{subfigure}{0.43\textwidth}
    \centering
    \includegraphics[width=\linewidth]{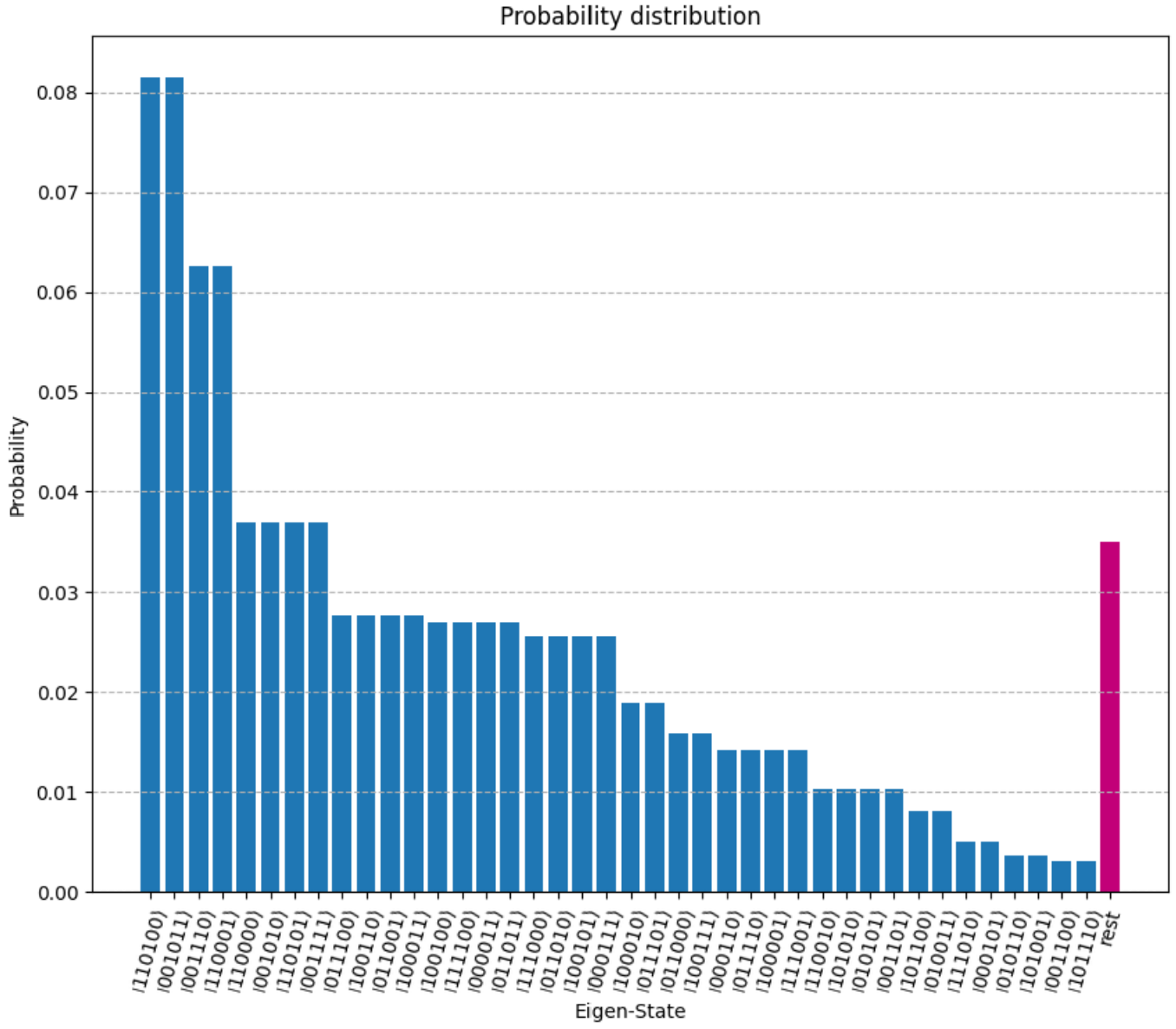}
    \captionsetup{justification=centering}
    \caption{Probability Distribution of Bitstrings}
    
  \end{subfigure}\hspace{-0.1cm}
  \begin{subfigure}{0.47\textwidth}
    \centering
    \includegraphics[width=\linewidth]{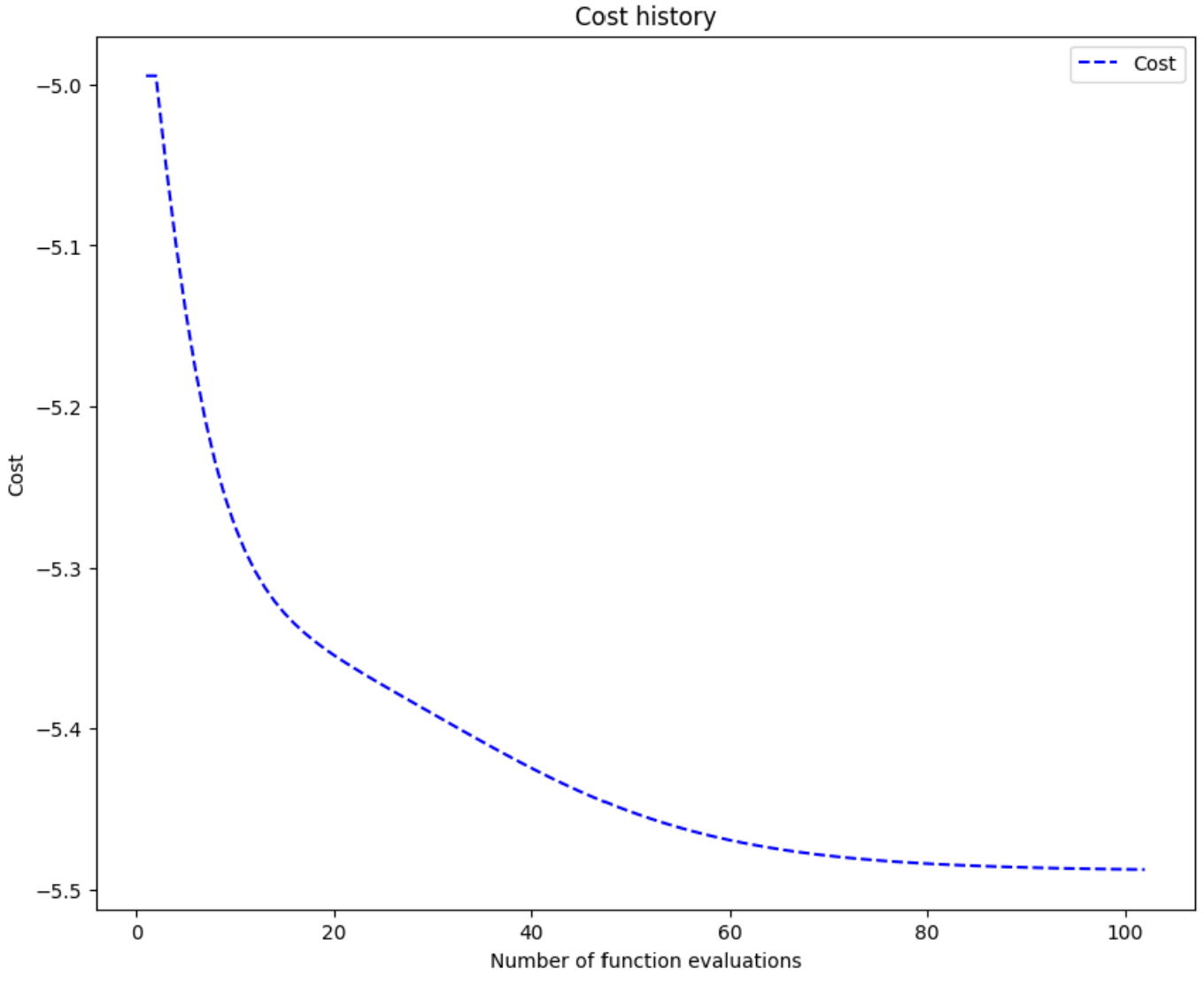}
    \caption{Classical Optimization Plot}
    
  \end{subfigure}
  %\vspace{-1cm}
  \caption{Visualization of outputs. The histogram on the left shows the probability distribution over bitstrings measured at the end of the quantum algorithm. Each bar corresponds to a particular bitstring, and the height indicates the probability of measuring that string. The line plot on the rights shows the classical optimization landscape, depicting the cost or objective value over iterations or parameter values. It provides insight into the performance of the classical optimization loop, and how well it’s navigating toward the minimum. }
  \label{fig:maxcut-plots}
\end{figure}

\begin{figure}[h]
  \centering
\includegraphics[width=0.45\textwidth]{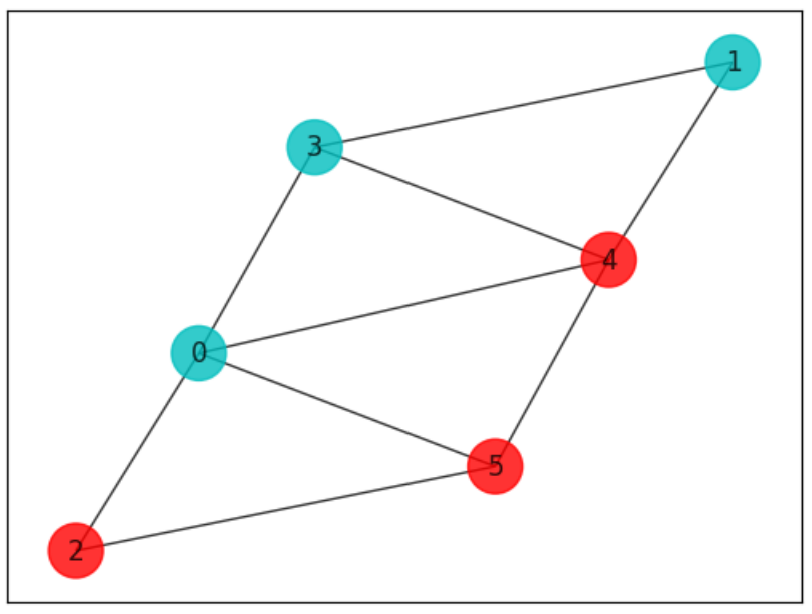}
  \caption{Optimal Graph Partitioning (Max Cut) based on Most Common Bitstring: (Nodes 2, 4, and 5) and Nodes (0, 1, 3).}
  \label{fig:example10}
  \label{fig:maxcut-sol}
\end{figure}

%An additional benefit of the OpenQAOA SDK is the easy modelling capabilities.
Once the circuit is created and executed, various visualization tools (Figure \ref{fig:maxcut-plots}) can be immediately implemented. These include a histogram of bitstring outcomes and their frequencies and the classical optimization plot. After postprocessing the most probable bitstring output, we obtain the optimal partition graph for the Maximum Cut (Figure \ref{fig:maxcut-sol}).

\subsection{QAOA via Qiskit}
\label{subsec:qqaoa}
Qiskit Optimization is a part of IBM Qiskit's open-source quantum computing framework. It provides a wide array of algorithms and built-in application classes for many canonical problems. 
%For this tutorial, we ignore the built-in application classes to take advantage of its Docplex modeling capabilities. By defining custom QUBOs instead of relying on predefined classes, users can more easily implement and solve their own QUBO models using Qiskit Optimization.
Qiskit Optimization is an extremely useful and easy way to solve QUBOs with various quantum algorithms including Variational Quantum Eigensolver \citep{Peruzzo_2014}, Adaptive Grover \citep{Gilliam_2021}, and QAOA. Furthermore, because of compatibility with IBMQ, algorithms can run on actual IBM quantum backends. In the following examples, all the code is run by default on the Qiskit QASM simulator. Although base-level implementation is easier, customization and in-depth analysis are more difficult. Tasks such as plotting optimization history or bit-string distributions, although possible, are significantly more challenging. The key limitations of Qiskit Optimization are its extremely long optimization times and a limited set of optimizers.  %It is extremely slow in solving QUBOs with over 10 variables.

%Although there is a pretty expansive set of local and global optimizers it is impossible to use any optimizer not listed here. Thus it is impossible to implement many promising algorithms like Adagrad and Genetic Algorithm as classical optimizers. Using Qiskit Runtime further customization such as error mitigation, custom ansatz circuits, and custom cost Hamiltonians are possible, but require more advanced knowledge. 

 We create a problem instance for Minimum Vertex Cover problem and its corresponding QUBO model.

\begin{lstlisting}[style=mystyle]
# Generating a graph of 4 nodes

n = 5  # Number of nodes in graph
G = nx.Graph()
G.add_nodes_from(np.arange(0, n, 1))
edges = [(0, 2, 1.0), (2, 4, 1.0), (1, 4, 1.0), (3, 4, 1.0)]
# tuple is (i,j,weight) where (i,j) is the edge
G.add_weighted_edges_from(edges)

colors = ["r" for node in G.nodes()]
pos = nx.spring_layout(G)

draw_graph(G, colors, pos)
\end{lstlisting}
\begin{figure}[h]
    \centering
    \includegraphics[width=0.6\linewidth]{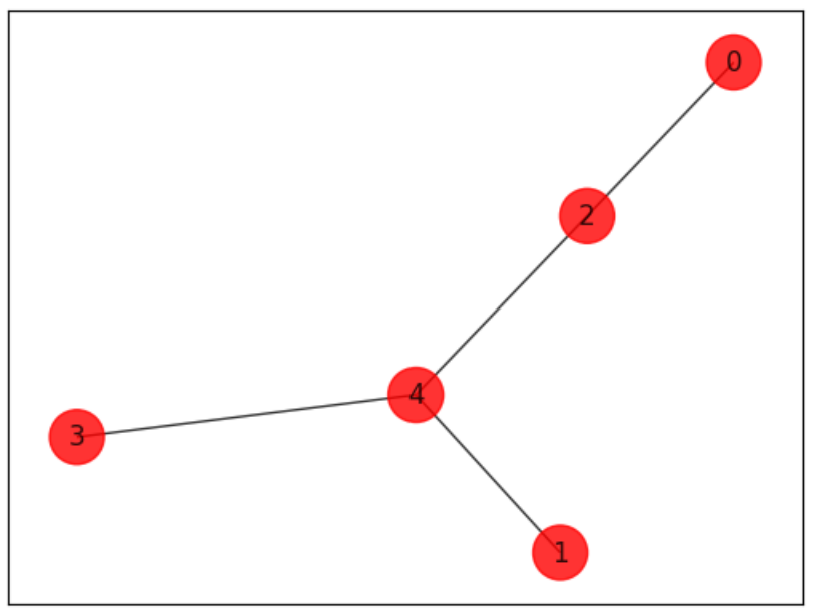}
\end{figure}
\begin{figure}[h]
  \centering
\includegraphics[width=0.6\textwidth]{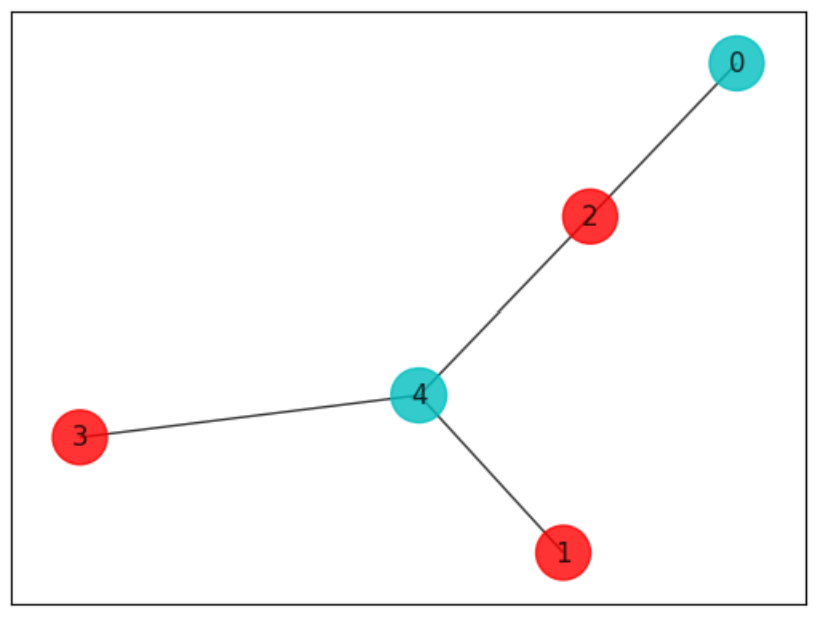}
  \caption{Optimal Minimum Vertex Cover: Nodes 0 and 4.}
  \label{fig:example15}
\end{figure}

\begin{lstlisting}[style=mystyle]
model = Model()
x = model.binary_var_list(n)
P = 10
H = sum(x[i] for i in range(n)) + P*sum(1 - x[e[0]] - x[e[1]] + x[e[0]]*x[e[1]] for e in edges)
model.minimize(H)
problem = from_docplex_mp(model)
qubo = QuadraticProgramToQubo().convert(problem)
\end{lstlisting}
Similarly to the Vanilla and OpenQAOA implementations, the following steps are to set up the QAOA circuit and optimize it. Because of the built-in capabilities of the Qiskit Optimization, this requires very little extra effort. Similarly to OpenQAOA, we just need to define a QAOA object. This object takes a sampler, optimizer, number of repetitions, and its properties are harder to modify.
\begin{lstlisting}[style=mystyle]
nelder_mead = NELDER_MEAD(maxiter=250)
sampler = Sampler()
qaoa = QAOA(sampler=sampler, optimizer=nelder_mead, reps=2)
algorithm = MinimumEigenOptimizer(qaoa)
\end{lstlisting}
We obtain the following solution to the Minimum Vertex Cover instance (Figure \ref{fig:example15}).
\subsection{Simulated Annealing}
\label{subsec:sai}
Let us now turn to the simulated annealing, which is used to solve the Order Partitioning Problem. Like the other algorithms, the first step is to create the problem instance and its corresponding QUBO model. As the Order Partitioning problem is custom to this paper, there are no pre-existing datasets or problem instances to use. Thus, we create a problem instance as shown below:
\begin{lstlisting}[style=mystyle]
Stocks = ['A', 'B', 'C', 'D', 'E', 'F']
stock_vals = [300, 100, 100, 200, 200, 100]
risk_factor_matrix = [[0.3, 0.1, 0.1, 0.2, 0.2, 0.1],
                      [0.4, 0.05, 0.05, 0.12, 0.08, 0.3],
                      [0.1, 0.2, 0.2, 0.3, 0.05, 0.05]]
T = sum(stock_vals)
n = 6 # number of stocks
m = 3 # number of risk factors
\end{lstlisting}
\begin{lstlisting}[style=mystyle]
x = Array.create('x', n, 'BINARY')
H1 = (T - 2*sum(stock_vals[j]*x[j] for j in range(n)))**2
H2 = sum(sum(risk_factor_matrix[i][j]*(2*x[j]-1)**2 for j in range(n)) for i in range(m))

# Construct Hamiltonian
a = Placeholder("a")
b = Placeholder("b")
H = a*H1 + b*H2
model = H.compile()

# Generate QUBO
feed_dict = {'a': 2, 'b': 2}
bqm = model.to_bqm(feed_dict=feed_dict)
\end{lstlisting}
Once the Order Partitioning binary quadratic model (BQM) is created, it is then run on the simulated annealing sampler. Additional parameters can be modified in the simulated annealing sampler such as the beta range, beta schedule, number of sweeps per beta, etc., where beta is the inverse temperature. For simplicity, we use the default parameters and read from the sampler $10$ times. Fewer reads are required than with quantum annealing, since the algorithm converges to an equilibrium more efficiently (although potentially not optimal). Like quantum annealing, the solutions with the lowest energy correspond to optimal outcomes.
\begin{lstlisting}[style=mystyle]
# Getting Results from Sampler
sa = neal.SimulatedAnnealingSampler()
sampleset = sa.sample(bqm, num_reads = 10)
decoded_samples = model.decode_sampleset(sampleset, feed_dict=feed_dict)
sample = min(decoded_samples, key=lambda x: x.energy)
\end{lstlisting}
The binary solution output from simulated annealing still needs to be sorted and paired with the corresponding stock. This is done explicitly in the associated Github repository. After this post-processing, we obtain the following solution:
\begin{figure}[h]
    \centering
    \includegraphics[width=0.8\textwidth]{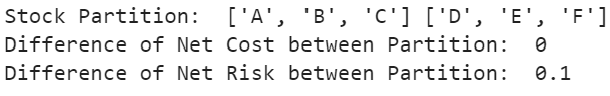}
    \caption{Solution to Order Partitioning instance.}
    \label{fig:enter-label15}
\end{figure}

\subsection{Quantum Annealing}
\label{subsec:qai}
The final algorithm we explore is quantum annealing, implemented using D-Wave's Quantum Annealer for the cancer genomic application. With access to significantly more qubits (a few thousand compared to IBM's 127 qubits), D-Wave enables the execution of larger problem instances.
%If run using the QAOA implementations discussed before, the problems would be solved far less efficiently in terms of both time and accuracy.

The cancer genomic problem requires significant preprocessing effort. First, to construct the QUBO cancer pathway, disease data must be read from some external source. For this tutorial, data were collected from The Cancer Genome Atlas Acute Myeloid Leukemia dataset posted on cBioPortal \citep{Ley2013GenomicAE}.

\begin{lstlisting}[style=mystyle]
# import necessary package for data imports
from bravado.client import SwaggerClient
from itertools import combinations

# connects to cbioportal to access data
cbioportal = SwaggerClient.from_url('https://www.cbioportal.org/api/v2/api-docs', config={"validate_requests":False,"validate_responses":False,"validate_swagger_spec":False})

# accesses cbioportal's AML study data
aml = cbioportal.Cancer_Types.getCancerTypeUsingGET(cancerTypeId='aml').result()

# access the patient data of AML study
patients = cbioportal.Patients.getAllPatientsInStudyUsingGET(studyId='laml_tcga').result()

# for each mutation, creates a list of properties associated with the mutation include geneID, patientID, and more
InitialMutations = cbioportal.Mutations.getMutationsInMolecularProfileBySampleListIdUsingGET(
    molecularProfileId='laml_tcga_mutations',
    sampleListId='laml_tcga_all',
    projection='DETAILED'
).result()
\end{lstlisting}
\noindent Using the data, a Patient-Gene dictionary was constructed that maps each patient to the mutated genes they have. The first four entries in the dictionary are shown below:

\begin{figure}[h]
    \centering
    \includegraphics[width=0.5\textwidth]{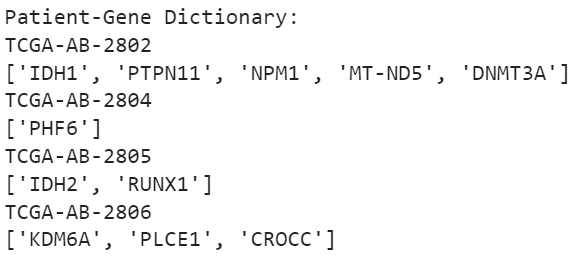}
    \caption{Patient-Gene Dictionary sample}
    \label{fig:enter-label2}
\end{figure}

Note that with this dictionary, both patients and genes are listed. The dictionary is used to create a diagonal matrix $\mathbf{D}$, where the index $d_{ii}$ of $\mathbf{D}$ represents the number of instances of gene $i$ amongst all patients.

\vspace{0.1in}

\begin{lstlisting}[style=mystyle]
# Initialize diagonal matrix
D = np.zeros((n, n))

# Populate diagonal matrix
for i, gene in enumerate(geneList):
   D[i, i] = sum(gene in genes for genes in PatientGeneDict.values())
\end{lstlisting}
%\vspace{-0.4cm}
\noindent To create the weighted adjacency matrix $\mathbf{A}_{w}$, the Patient-Gene dictionary is modified to contain Gene Pairs. For example, if a patient possesses gene $i$ and gene $j$, they possess the gene pair $(i, j)$. We first create a helper function to create all pairs of entries in a list and then use a dictionary to map each patient to its list of gene pairs. This new patient-gene pair dictionary is used to create matrix $\mathbf{A}$, where the index $A_{ij}$ represents the number of patients who possess gene pair $(i, j)$. By symmetry, this quantity is also stored at $A_{ji}$.

\begin{lstlisting}[style=mystyle]
# helper function to generate all gene pairs from a list of genes
def generate_pairs(list):
    pairs = set()
    for subset in combinations(list, 2):
        pairs.add(tuple(sorted(subset)))
    return pairs
\end{lstlisting}
\vspace{-0.4cm}
\begin{lstlisting}[style=mystyle]
# creates a patient-(gene-list-pair) dictionary
PatientGeneDictPairs = {}
for m in mutations:
    PatientGeneDictPairs[m.patientId] = generate_pairs(PatientGeneDict[m.patientId])
\end{lstlisting}
\vspace{-0.4cm}
\begin{lstlisting}[style=mystyle]
A = np.zeros((n, n))

for i in range(n):
    for j in range(i + 1, n):  # iterate over upper triangular part
        count = sum((geneList[i], geneList[j]) in pairs for pairs in PatientGeneDictPairs.values())
        A[i, j] = A[j, i] = count  # Exploit symmetry
\end{lstlisting}
\vspace{-0.4cm}
\noindent Once the $\mathbf{D}$ and $\mathbf{A}$ matrices are constructed, we create the QUBO model and later the BQM to run on D-Wave. This BQM is unconstrained, as it uses unconstrained binary variables to model the pathway problem. %There are other models, like Constrained Quadratic Models and Discrete Quadratic Models, which are applied to other types of problems, but the Binary Quadratic Models are the best for decision making optimization problems like the Cancer Genomic problem.

\begin{lstlisting}[style=mystyle]
# initializes an array of QUBO variables
x = Array.create('x', n, 'BINARY')
H1 = sum(sum(A[i][j]*x[i]*x[j] for j in range(n)) for i in range(n))
H2 = sum(D[i][i]*x[i] for i in range(n))
a = Placeholder("alpha")
H = H1 - a*H2

# creates a Hamiltonian to run on D-Wave sampler
model = H.compile()
feed_dict = {'alpha': 0.45}
bqm = model.to_bqm(feed_dict=feed_dict)
\end{lstlisting}
Once the BQM is created, it is then mapped to the qubits on the Quantum Processing Unit (QPU). The sampler does this process by encoding the problem into physical qubits and couplers. Recall the original form of a QUBO model:
\[\sum_{1 \leq j < i \leq n} Q_{ij}x_{i}x_{j} + \sum_{i=1}^{n}Q_{ii}x_{i}.\]
Embedding is the process of mapping the problem to a graph where $Q_{ii}$ is the bias of the nodes (the physical qubits) and $Q_{ij}$ is the weight of the edges (the couplers). This mapping from the binary quadratic model to the QPU graph is known as a minor embedding; however, other types of embedding do exist for different problems. Composites allow for higher-level customization of the annealing process options such as parallel QPU processing, automatic embedding, etc. The Ocean API is very powerful and provides a plethora of customizable features for samplers, embeddings, and composites, but for the sake of simplicity, we use the standard D-Wave Sampler and Embedding Composite.
\begin{lstlisting}[style=mystyle]
# Getting Results from Sampler
sampler = EmbeddingComposite(D-WaveSampler())
sampleset = sampler.sample(bqm, num_reads=1000)
sample = sampleset.first
\end{lstlisting}
The annealing process will return for each read, so we call ``sample.first" to return the one with the lowest energy. Typically, variables are unsorted, so they must be sorted to correspond correctly to the genes. This post-processing is tedious and is included in the associated Github, and provides cancer genome pathways (one example shown below in Figure \ref{fig:enter-label10}). We also calculate the properties of the pathways (coverage, coverage/gene, independence, measure).
\begin{figure}[h]
    \centering
    \includegraphics[width=0.5\textwidth]{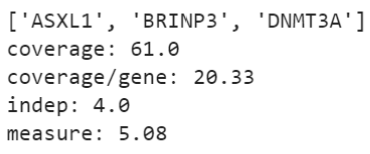}
    \caption{Cancer Gene Pathway discovered through quantum annealing.}
    \label{fig:enter-label10}
\end{figure}

\section{Concluding Remarks and Future Directions}
\label{sec:conclusion}
The purpose of this tutorial is to rapidly introduce how to (a) represent canonical and practical problems as Quadratic Unconstrained Binary Optimization (QUBO) models and (b) solve them using (i) gate-circuit quantum computers (such as IBM) or their simulators using the Quantum Approximate Optimization Algorithm (QAOA), (ii) quantum annealing (on D-Wave), and (iii) simulated annealing (classical). We illustrate the above algorithms on three canonical problems and two practical problems (one from cancer genomics and the other from hedge fund portfolio management). A summary of the various algorithms we have discussed is presented in Table~\ref{table:implementation_comparison}.

To fully exploit the capabilities of nascent quantum devices, it is imperative to advance decomposition methods that enable the effective partitioning of large quantum circuits into smaller hardware-compatible components. This remains an active research frontier, with recent efforts exploring tensor network decompositions \citep{Ran2020}, circuit cutting techniques \citep{Bravyi2020}, and modular algorithm designs. Such approaches are essential to overcome current limitations in the qubit counts, the coherence times, and the connectivity. While waiting for hardware to become mature, researchers are also developing quantum-inspired methods \citep{TT24}, building photonic devices such as Ising solvers \citep{PS2023}, or novel decompositions based on Algebraic Geometry and solving embedded QUBO sub-problems using SA \citep{GP2024, KT2024}.

Beyond decomposition, significant future work is needed in developing robust error mitigation strategies to counteract the noise and decoherence effects prevalent in near-term quantum hardware \citep{Temme2017, Kandala2019}. Furthermore, hybrid quantum-classical algorithms, including variational methods \citep{mcclean2016theory} and quantum machine learning models \citep{Biamonte2017}, require further optimization and scaling to demonstrate practical advantage.

Hardware-aware compilation and algorithm optimization represent another critical research avenue, with the aim of tailoring quantum circuits to specific device architectures to achieve enhanced performance and reduced error rates \citep{JavadiAbhari2020, Shafaei2021}. Moreover, standardized benchmarking and validation protocols must be refined to reliably assess device capabilities and algorithm efficacy \citep{Proctor2022}.

Collectively, these research directions will be fundamental to bridging the gap between theoretical quantum algorithms and their practical implementation on emerging quantum platforms, ultimately accelerating progress toward scalable and fault-tolerant quantum computing. Interested readers can learn more about the state of quantum algorithms and quantum-inspired algorithms for optimization applications in \citep{Gomes2024}.
%\vspace{-0.9cm}
\begin{table}[h!]
\centering
\renewcommand{\arraystretch}{1.5} % Adjust row height for readability
\begin{tabular}{| p{4cm} | p{10cm} |}
\hline
\textbf{Implementation} & \textbf{Characteristics} \\ 
\hline
\textbf{QAOA (Vanilla)} & 
\begin{itemize}
    \item High circuit customization
    \item High optimizer customization
    \item Difficult implementation
    \item Accuracy substantially decreases with problem size
\end{itemize} \\ 
\hline
\textbf{QAOA (OpenQAOA)} & 
\begin{itemize}
    \item Moderate circuit customization
    \item Moderate optimizer customization
    \item Easy implementation
    \item Best time vs. accuracy tradeoff among implementations
\end{itemize} \\ 
\hline
\textbf{QAOA (Qiskit)} & 
\begin{itemize}
    \item Low circuit customization
    \item Low optimizer customization
    \item Easy implementation
    \item Time exponentially increases with problem size
\end{itemize} \\ 
\hline
\textbf{Simulated Annealing} & 
\begin{itemize}
    \item High parameter customization
    \item Handles larger problem sizes
    \item Fastest runtime
    \item Most accurate results
    \item Easy implementation
\end{itemize} \\ 
\hline
\textbf{Quantum Annealing} & 
\begin{itemize}
    \item High topology customization (not discussed in this paper)
    \item Handles larger problem sizes
    \item Fastest quantum runtime
    \item Most accurate quantum results
    \item Easy implementation
\end{itemize} \\ 
\hline
\end{tabular}
\caption{Comparison of Various Implementations and Their Characteristics.}
\label{table:implementation_comparison}
\end{table}

%\noindent {\bf Acknowledgements.} The authors thank Claudio Gomes, Anil Prabhakar, Elias Towe, Shrinath Viswanathan, and Daniel Mai for their comments on the previous version.

\bibliographystyle{informs2014}  
{\small
\bibliography{main}}

\end{document}